\documentclass[11pt,a4paper]{article}
\pdfoutput=1

\usepackage[utf8]{inputenc}
\usepackage{jheppub}
\usepackage{latexsym}
\usepackage{color}
\usepackage[usenames,dvipsnames,svgnames,table]{xcolor}
\usepackage{tikz}
\usepackage{graphicx, braket}
\usepackage{epsfig}  
\usepackage{epsf}   
\usepackage{float}
\usepackage[normalem]{ulem}
\usepackage[normalem]{ulem}
\usepackage{dcolumn}
\usepackage{bm}
\usepackage[toc,page]{appendix}
\usepackage{dcolumn}
\usepackage{textcomp}
\usepackage{float}
\usepackage{hypcap}
\usepackage[]{hyperref}
\usepackage{booktabs}
\usepackage{multirow}
\usepackage{diagbox}
\usepackage{alphalph}
\usepackage{textcomp}
\usepackage{amsmath}
\usepackage{epstopdf}
\usepackage{orcidlink}
\usepackage{url}

\usepackage[utf8]{inputenc}

\usepackage{lineno}
\usepackage{soul} 
\usepackage[normalem]{ulem}
\usepackage{longtable}
\usepackage{multirow}
\usetikzlibrary{svg.path}

\definecolor{lime}{HTML}{A6CE39}
\DeclareRobustCommand{\orcidicon}{\hspace{-2.1mm}
\begin{tikzpicture}
\draw[lime,fill=lime] (0,0.0) circle [radius=0.13] node[white] {{\fontfamily{qag}\selectfont \tiny ID}}; \draw[white,fill=white] (-0.0525,0.095) circle [radius=0.007]; 
\end{tikzpicture} \hspace{-3.5mm} }
\foreach \x in {A, ..., Z} {\expandafter\xdef\csname orcid\x\endcsname{\noexpand\href{https://orcid.org/\csname orcidauthor\x\endcsname} {\noexpand\orcidicon}}}

\newcommand{\beq}{\begin{equation}}
\newcommand{\eeq}{\end{equation}}
\newcommand{\beqa}{\begin{eqnarray}}
\newcommand{\eeqa}{\end{eqnarray}}
\definecolor{orcidlogocol}{HTML}{A6CE39}

\makeatletter
\gdef\@fpheader{}
\makeatother

\preprint{IP/BBSR/2026-09}

\title{Quantum Information as a New Lens for Precision Neutrino Physics}

\author[a]{Khushboo Dixit\,\orcidA}
\author[b,c]{Ritam Kundu\,\orcidB}
\author[d]{Papia Panda\,\orcidC}
\author[a,e,f]{Soebur Razzaque\,\orcidD}
\author[b,c,g]{Ramita Sarkar\,\orcidE}

\affiliation[a]{Centre for Astro-Particle Physics (CAPP) and Department of Physics, University of Johannesburg, PO Box 524, Auckland Park 2006, South Africa}
\affiliation[b]{Institute of Physics, Sachivalaya Marg, Sainik School Post, Bhubaneswar 751005, India}
\affiliation[c]{Homi Bhabha National Institute, Training School Complex, Anushakti Nagar, Mumbai 400094}
\affiliation[d]{School of Physics, University of Hyderabad, Hyderabad - 500046, India}
\affiliation[e]{Department of Physics, The George Washington University,
Washington, DC 20052, USA}
\affiliation[f]{National Institute for Theoretical and Computational Sciences (NITheCS),
Private Bag X1, Matieland, South Africa}
\affiliation[g]{Department of Chemical Engineering, Engineering Faculty, Ariel University, Ariel 407000, Israel}

\emailAdd{kdixit@uj.ac.za}
\emailAdd{kunduritam14@gmail.com}
\emailAdd{ppapia93@gmail.com}
\emailAdd{srazzaque@uj.ac.za}
\emailAdd{ramitasarkar11@gmail.com}

\abstract{We present a quantum-information-theoretic study of three-flavor neutrino oscillations in long-baseline experiments by mapping flavor states to qubit-like representations and quantifying quantum correlations through total concurrence. The local minima of this entanglement measure identify energy regions where the flavor state is closest to separability, enabling cleaner extraction of oscillation parameters. We explain how these local minima offer opportunities for precision measurements and provide insight into the accurate determination of neutrino oscillation parameters. We then propose a strategy to improve parameter extraction by aligning the benchmark oscillation regions of NO$\nu$A and T2K with the minimum entanglement achievable in each experiment. This shifts the concurrence minima toward higher-event-count energy regions, leading to tighter constraints and reducing the tension arising from their different energy regimes. For normal ordering, we obtain $(0.581^{+0.0136}_{-0.0150},195^{+38}_{-32}\,^\circ)$ in the $(\sin^2\theta_{23},\delta_{\rm CP})$ plane and $(0.580^{+0.0140}_{-0.0153},2.515^{+0.0344}_{-0.0344}\times10^{-3}\,\mathrm{eV}^2)$ in the $(\sin^2\theta_{23},\Delta m^2_{31})$ plane, yielding improved joint constraints. Using GLoBES simulations together with real data, we assess how local minima of quantum correlations influence leptonic CP-violation sensitivity, $\theta_{23}$ octant-degeneracy resolution, and mass-ordering determination. Our results show that minimizing entanglement can significantly affect these key sensitivities, highlighting quantum information measures as complementary probes of neutrino flavor oscillations and offering new insight into the role of quantum correlations in precision neutrino physics.
}


\arxivnumber{ 2606.31996 }

\begin{document}
\maketitle

\section{Introduction}
Quantum information and entanglement studies have emerged as one of the cornerstones driving several frontiers of particle physics over the past few decades. From experiments at the Large Hadron Collider (LHC)~\cite{Barr:2024djo,Zhang:2025mmm,ATLAS:2023fsd} to astroparticle observations~\cite{Pfenniger:2006rd, Dixit:2019lsg, Balantekin:2022lys, Roggero:2022hpy, Dixit:2024pcj, SinghKoranga:2026kiu, Banerjee:2026jpv} and neutrino phenomenology~\cite{Dixit:2021jrk, Dixit:2024mbf}, the study of quantum correlations, including entanglement, has demonstrated significant potential, opening new avenues for exploration. The interdisciplinary interface between two rapidly evolving fields, quantum information and neutrino oscillation physics has now become a topic of considerable interest, even without invoking exotic right-handed neutrinos or extending the Standard Model (SM) with additional symmetries. A key conceptual development in quantum information science is the realization of quantum entanglement within a single particle among its different modes, which has provided a novel framework to understand the nature of entanglement both qualitatively and quantitatively~{\color{red}{\cite{vanEnk:2005bhd, Zanardi:2002pwt, Shi:2003zz}}}. This framework has been substantiated through experimental observations in single-photon systems~\cite{PhysRevLett.92.180401, PhysRevLett.88.070402, PhysRevLett.87.050402}. In the context of neutrino physics, the idea of entanglement for a single neutrino distributed among different flavor modes has long been a subject of interest within the community, in the backdrop of two-flavor oscillation \cite{Blasone:2007vw, Burrage:2025xac}. Subsequently, with the establishment of the $3\nu$ framework through results from experiments such as MINOS+~\cite{Evans:2017brt, MINOS:2020llm}, NO$\nu$A~\cite{NOvA:2007rmc, NOvA:2021nfi}, and T2K~\cite{T2K:2023smv}, along with the discovery of non-zero $\theta_{13}$ by Daya Bay reactor neutrino experiment~\cite{DayaBay:2012fng, DayaBay:2022orm}, quantum correlation studies in the neutrino sector have gained significant momentum~\cite{Blasone:2007vw, Banerjee:2015mha, Formaggio:2016cuh, Dixit:2018gjc, Ettefaghi:2020otb, Li:2022mus, Alam:2026bxn, Bouri:2024kcl}.

The quantum nature of neutrino oscillations arises from the superpositions of mass eigenstates. When these states are expressed in the flavor basis, they exhibit correlations between different flavor modes. In this notion, we can quantify these correlations through entanglement using well-established measures such as concurrence~\cite{Zanardi:2002pwt, Schliemann:2001rtw}. These quantities offer an alternative characterization of flavor transition phenomena beyond conventional probability-based analyses. Importantly, entanglement measures depend on the oscillation parameters, including the mixing angles, mass-squared splittings, and the Dirac CP phase, and may therefore help in resolving parameter degeneracies and probing leptonic CP violation~\cite{Blasone:2007vw, Banerjee:2015mha}. Moreover, a three-flavor entanglement structure may also indicate the presence of new physics effects, such as non-standard interactions~\cite{Dixit:2019swl,Sarkar:2020vob, Konwar:2025ipv}, decoherence~\cite{Dixit:2018gjc}, or additional sterile degrees of freedom~\cite{Benatti:2000ph, Fiza:2021gvq}.

Long-baseline neutrino experiments are featured by relatively lower uncertainties in the beam flux and improved control over the oscillation length and neutrino energy compared to other neutrino experiments. Hence, even subtle departures from the standard entanglement structure are expected to leave observable imprints on the sensitivities of various outcomes driven by oscillation phenomenology in these unprecedented long-baseline setups. 
Motivated by the significant role of NO$\nu$A and T2K experiments in shaping the global fits of neutrino oscillation data on behalf of the long-baseline sector, we investigate in this work how concurrence, a measure of quantum entanglement can influence some of the long-standing open issues in the standard $3\nu$ framework. In our analysis, a single neutrino is treated as a generalized tripartite system comprising three flavor modes ($\nu_e,\,\nu_\mu,\,\text{and}\,\,\nu_\tau$), which can be mutually entangled. We consider the total concurrence as a measure of quantum entanglement to quantify the degree of entanglement among different flavor modes. It is an important fact to mention that, in our work, we do not consider the interactions between two neutrinos or possible entanglements between them, treating them as individual and independent particles. Following the footprints of the seminal work of the concurrence of an arbitrary state of two qubits~\cite{Wootters:1997id}, we map the idea of concurrence on the canvas of $3\nu$ framework. The central theme of this work is to probe the impact of the extremal realizations, specifically the effect of local minima, of this entanglement measure on neutrino oscillation parameters, with particular emphasis on leptonic CP violation, the resolution of the $\theta_{23}$ octant ambiguity, mass ordering determination, and the enhancement or deterioration of the bounds in reshaping of the allowed parameter space in the $(\sin^2\theta_{23} - \delta_\mathrm{CP})$ and $(\sin^2\theta_{23} - \Delta m_{31}^2)$ planes. Through this investigation, we aim to understand, within the oscillation canvas and for a single neutrino system, how the limiting scenario of complete separability among flavor modes affects the sensitivities to CP violation, rejection of the incorrect $\theta_{23}$ octant, and the mass ordering determination, in the backdrop of the current data from the ongoing NO$\nu$A and T2K experiments.

NO$\nu$A and T2K are two currently running long-baseline experiments poised to determine the neutrino mass ordering and measure $\theta_{23}$ and CP-phase ($\delta_{\rm CP}$). Results from the two experiments, however, differ significantly on their individual best-fit value of $\delta_{\rm CP}$~\cite{Cherchiglia:2023ojf, Rahaman:2022rfp, Rahaman:2021zzm, Chatterjee:2024kbn, Nizam:2018got, Prakash:2013tta}. In addition to several complementary features, a key distinction between T2K and NO$\nu$A lies in their different operating energy regimes and baselines, which lead them to probe the same flavor state under substantially different oscillation conditions: T2K has a flux peak around 0.6 GeV at a baseline of 295 km, whereas NO$\nu$A has a flux peak around 2 GeV at a baseline of 810 km. Given the energy and baseline dependence of the total concurrence through the oscillation probabilities, our prescription of choosing the common oscillation parameters to optimize the entanglement associated with the flavor state for both experiments places T2K and NO$\nu$A on a common footing, irrespective of their different operational energy and baseline. This method avoids anchoring our analysis to the individual best-fit values of either experiment or to a particular global best-fit benchmark. \\

The structure of this paper is organized as follows: we present the theoretical framework in Section~\ref{sec:TheoreyFramework}, where we define total concurrence in the context of three-flavor neutrino oscillations. In Section~\ref{sec:ExperimentalDetails}, we explore the specifics of the experimental scenarios for T2K and NO$\nu$A, followed by a discussion on how to resolve the tensions between these two experiments from the perspective of total concurrence. Finally, we present our results in Section~\ref{sec:Results}, followed by discussion and conclusions in Section~\ref{sec:Discussion}.

\label{sec:introducion}
  

\section{Theoretical Framework}
\label{sec:TheoreyFramework}
In this section, we start our discussion by introducing the Pontecorvo-Maki-Nakagawa-Sakata (PMNS) parameterization in the standard three-flavor scenario~\cite{Maki:1962mu,Pontecorvo:1967fh}. Further, we provide the theoretical framework for constructing the total concurrence, the entanglement measure to quantify the entanglement distributed among its flavor modes, for the three-flavor neutrino oscillation scenario by mapping the neutrino flavor state onto three-qubit states. 

\subsection{$3\nu$ flavor oscillations}
Neutrino oscillation is a robust phenomenon in particle physics that gave the first signature of Beyond Standard Model (BSM) physics. This phenomenon confirms the existence of non-zero neutrino mass. The evidence of non-zero $\theta_{13}$ from the Daya Bay experiment~\cite{DayaBay:2012fng, DayaBay:2022orm} firmly established the three-flavor neutrino framework. Neutrinos are produced and detected in their flavor states (\textit{i.e.,} $\nu_e$, $\nu_\mu$, or $\nu_\tau$). However, during propagation through vacuum or matter, they evolve as mass eigenstates (\textit{i.e.,} $\nu_1$, $\nu_2$, and $\nu_3$), which are the eigenstates of the Hamiltonian. Neutrino oscillation implies that there is a finite probability that a neutrino is observed in a flavor different from its initial one at the detector, and this probability depends on the baseline length and the neutrino energy. These two sets of states (flavor states and mass eigenstates) are connected by a unitary matrix to conserve the total oscillation probability. This unitary matrix contains three mixing angles: atmospheric mixing angle $\theta_{23}$, reactor mixing angle $\theta_{13}$, and solar mixing angle $\theta_{12}$; and a Dirac CP phase $\delta_\mathrm{CP}$, considering neutrino as a Dirac particle. Each of the mass eigenstates can be expressed as a coherent and linear superposition of the flavor states and vice versa. The amount of the superposition of a particular flavor to a given mass eigenstate is governed by the oscillation parameters. Mathematically, it can be written as, 
\begin{equation}
\ket{\nu_{\beta}} = \sum^3_{l=1} U_{\beta l} \, \ket{\nu_l},
\label{eqn:relation_flavor_mass}
\end{equation}
at $t=0$. Here, $\beta$ and $l$ denote the flavor ($e,\,\mu,\,$ $\tau$) and mass indices (1, 2, 3), respectively and $U_{\beta l}$ are the elements of the PMNS matrix,
\begin{align}
U = & \begin{pmatrix}
c_{12}c_{13} & s_{12}c_{13} & s_{13}e^{-i\delta_\mathrm{CP}} \\
-s_{12}c_{23}-c_{12}s_{23}s_{13}e^{i\delta_\mathrm{CP}} & c_{12}c_{23}-s_{12}s_{13}s_{23}e^{i\delta_\mathrm{CP}} & c_{13}s_{23} \\
s_{12}s_{23}-c_{12}s_{13}c_{23}e^{i\delta_\mathrm{CP}} & -c_{12}s_{23}-s_{12}s_{13}c_{23}e^{i\delta_\mathrm{CP}} & c_{13}c_{23}
\end{pmatrix}.
\label{eqn:PMNS_matrix}
\end{align}\\
Here, we take the convention, $c_{ij}=\cos\theta_{ij}$ and $s_{ij}=\sin\theta_{ij}$ $\forall$ $i<j=1,2,3$. Following this parameterization, the probability that a neutrino of flavor $\alpha$ oscillates to a new flavor $\beta$ after traveling a distance $L$ (km) and having energy $E$ (GeV) can be expressed~\cite{Giunti:2007ry} as follows, \\
\begin{align}
P_{\alpha \beta} &= \delta_{\alpha \beta}
- 4 \sum_{i>j} \mathrm{Re}\!\left( U_{\alpha i} U^*_{\beta i} U^*_{\alpha j} U_{\beta j} \right)
\sin^2\!\left( 1.27\, \frac{\Delta m^2_{ji} L}{E} \right) \nonumber \\
&\quad \pm 2 \sum_{i>j} \mathrm{Im}\!\left( U_{\alpha i} U^*_{\beta i} U^*_{\alpha j} U_{\beta j} \right)
\sin\!\left( 2.54\, \frac{\Delta m^2_{ji} L}{E} \right).
\end{align}\\
The positive (negative) sign of the third term, \textit{i.e.,} CP-violating term corresponds to oscillation of neutrino (antineutrino). Here, $\Delta m^2_{ji}=m^2_j-m^2_i$ (eV$^2$) denotes the solar or atmospheric mass-squared splitting that can be measured experimentally.
The governing Hamiltonian $\mathcal{H}$ in the flavor basis representing the kinetic term of neutrinos and the interaction with the Earth matter can be shown as,
\begin{equation}
\mathcal{H}= \frac{1}{2E}\bigg[U^\dagger \begin{pmatrix}
0 & 0 & 0 \\
0 & \Delta m^2_{21} & 0 \\
0 & 0 & \Delta m^2_{31}
\end{pmatrix} U\bigg] + \begin{pmatrix}
V_{CC} & 0 & 0 \\
0 & 0 & 0 \\
0 & 0 & 0
\end{pmatrix}.
\label{equn:Hamiltonian}
\end{equation}
Here, $U$ stands for the PMNS matrix as mentioned in eq.~\ref{eqn:PMNS_matrix}, $E$ represents the neutrino energy, $V_{CC}$ depicts the Earth matter potential due to charged current (CC) interaction of neutrinos with the Earth. Here, $V_{CC}= \pm \sqrt{2}G_F N_e = \pm (7.6\times 10^{-14}\times Y_e\times\rho)$ eV, where $G_F = 1.667\times10^{-5}\, \mathrm{GeV}^{-2}$ is the Fermi coupling constant, $N_e$ stands for the electron number density in the Earth matter, $\rho$ (in $g/cm^3$) is the line-averaged matter density of the medium, and $Y_e$ is the fractional contribution of electrons of the Earth (generally, upon the assumption of equal contribution of muon and tau in electrically neutral matter; $Y_e=\frac{N_e}{N_p+N_n}=0.5$ )~\cite{Wolfenstein:1977ue, Mikheyev:1985zog, Mikheyev1986, Minakata:2015gra, Denton:2016wmg}. The matter potential changes sign to negative for antineutrinos. Eq.~\ref{equn:Hamiltonian} can be simplified as, 
\begin{equation}
    \mathcal{H} = \frac{\Delta m_{31}^2}{2 E} \left[ R_{23} U_{\delta} M U_{\delta}^{\dagger} R_{23}^T \right],
\end{equation}
with $U_{\delta} = \mathrm{diag}(1,1,e^{i\delta_{CP}})$ and $M = [R_{13} R_{12},\mathrm{diag}(0,\alpha,1),R_{12}^{T} R_{13}^{T}]$, with $\alpha = \Delta m^2_{21}/\Delta m^2_{31}$. After diagonalizing the Hamiltonian $\mathcal{H}$, we obtain the energy eigenvalues $E_1, E_2$, and $E_3$, along with the corresponding eigenvectors $v_1, v_2$, and $v_3$. The resulting mixing matrix is given by $W=(v_1~~v_2~~v_3)$, and the effective mixing matrix in matter takes the form $\mathcal{U}=R_{23}U_{\delta}W$.

Neutrino mass-eigenstates are the stationary states, $i.e.,$ eigenstates of the free Hamiltonian (or interacting Hamiltonian in the presence of Earth matter potential). Hence, their time evolution can be represented as a solution to the Schrödinger equation, given as $\ket{\nu_l (t)} = e^{-i E_l t} \ket{\nu_l (0)}$, with $E_l$ being the energy of neutrinos of $l$-th mass eigenstate. Consequently, the connection between mass eigenstates and the flavor states given in eq.~\ref{eqn:relation_flavor_mass} will be modified at $T=t$ by the phase factor as follows,
\begin{equation}
\ket{\nu_{\beta} (T=t)} = \sum^3_{l=1} \mathcal{U}_{\beta l}~e^{-iE_lt} \, \ket{\nu_l (T=0)},
\label{eqn:flavortomass}
\end{equation}
where $\mathcal{U}_{\beta l}$'s are the component of the matter-modified mixing matrix $\mathcal{U}$. The connection between two flavor states at different time points, using eqs.~\ref{eqn:relation_flavor_mass} and \ref{eqn:flavortomass}, can be depicted as follows, 
\begin{align}
\ket{\nu_{\beta} (t)} 
&= \sum^3_{l=1} \sum_{\alpha=e,\mu,\tau}  \mathcal{U}_{\beta l}~e^{-iE_lt} ~\mathcal{U}^*_{\alpha l} \, \ket{\nu_\alpha (0)} \,\, 
\nonumber\\
& = \sum_{\alpha=e,\mu,\tau} \widetilde{\mathcal{U}}_{\beta\alpha} \ket{\nu_\alpha (0)},
\end{align}
where, $\widetilde{\mathcal{U}}_{\beta\alpha}= \sum^3_{l=1} \mathcal{U}_{\beta l}~e^{-iE_lt} ~ \mathcal{U}^*_{\alpha l}$, denotes the flavor transition amplitude, which is subsequently used to compute the quantum entanglement measure, concurrence.

\subsection{Measure of entanglement}
	Entanglement is a key property of composite quantum systems where the state cannot be described as a product of individual subsystems. In neutrino oscillations, the mass eigenstates form an entangled basis in the flavor state representation by means of coherent and linear superposition. In this study, we thoroughly examine the degree of entanglement for a single neutrino flavor state, employing total concurrence as our analytical framework.
\\

	Concurrence, being an important measure of entanglement~\cite{Hill:1997pfa,Wootters:1997id,nielsen_chuang_2010} is strictly positive for
	entangled states and vanishes for all separable states. Thus, 
concurrence provides a compact entanglement monotone that captures bipartite quantum 
correlations, satisfying necessary and sufficient conditions for separability. For a composite system with two sub-systems $M$ and $M^\prime$ in it, the concurrence associated with a bipartition $M|M^\prime$ is defined as~\cite{Hill:1997pfa},
	\begin{equation}
		C^2_{M| M ^\prime}=2[1-tr{(\rho^M)^2}],
	\end{equation}
	where $\rho^M$ is the reduced density matrix of the sub-system $M$. Thus, concurrence directly quantifies the loss of purity of a subsystem due to entanglement with the other subsystem. Equivalently, for a two-qubit (AB) state, concurrence can also be written in the form,
	\begin{equation}
		C^2_{A|B} = 4 ~\text{det}(\rho^A).
	\end{equation}  
The idea of concurrence has also been extended beyond two-qubit systems to quantify 
entanglement in multipartite states. Since multipartite entanglement is not captured 
by a single unique quantity, several generalizations of concurrence have been 
proposed. A common strategy is to construct a global or total concurrence by 
combining the concurrences associated with different bipartitions or different 
two-mode reductions of the multipartite state. In this sense, the total concurrence 
acts as a collective measure of how entanglement is distributed among the subsystems.
For a three party system consisting of sub-systems $A$, $B$, and $C$, there are three possible bipartitions: $A|BC$, $B|CA$, and $C|AB$. In this case, we define total entanglement as the sum of entanglement across all bipartitions, specifically in terms of the sum of all one-vs-rest squared concurrences,~\cite{Wootters:1997id, sarkar2021geometry, dutta2019permutation}
\begin{equation}
    C^2_{\mathrm{tot}} = C^2_{A|BC} + C^2_{B|CA} + C^2_{C|AB}.
\end{equation}
Here, each one-vs-rest squared concurrence, corresponding to one bipartition, say $A|BC$, follows the Coffman-Kundu-Wootters (CKW) monogamy relation~\cite{Hill:1997pfa, Coffman:1999jd} and can be decomposed into pairwise ($A|B$, $A|C$) concurrence contributions and the residual three-tangle depicting the genuine residual three-party contribution. However, for the single-excitation flavor-mode state relevant to neutrino oscillations, the residual three-tangle vanishes, so the measure captures the distributed pairwise flavor-mode entanglement.

	\subsection{Neutrino flavor state in occupation number representation}
	
	The evolved neutrino state with flavor $\beta$ in a coherent superposition of flavor basis can be written as,
	\[
	\ket{\nu_\beta(t)} = \widetilde{\mathcal{U}}_{\beta e}(t)\ket{\nu_e} + \widetilde{\mathcal{U}}_{\beta \mu}(t)\ket{\nu_\mu} + \widetilde{\mathcal{U}}_{\beta \tau}(t)\ket{\nu_\tau}.
	\] 
	In the effective two-flavor $(\nu_e - \nu_\mu)$ oscillation scenario, the flavor states can be mapped onto a two-qubit state using the occupation number representation at $t=0$ as,
	\[
	\ket{\nu_e} = \ket{1}_e \otimes \ket{0}_\mu \equiv \ket{10}_e, \\
	\ket{\nu_\mu} = \ket{0}_e \otimes \ket{1}_\mu \equiv \ket{01}_\mu,
	\]
	where, $\ket{1}_e= \begin{pmatrix}
		0\\
		1
	\end{pmatrix}$ and  $\ket{0}_\mu= \begin{pmatrix}
		1\\
		0
	\end{pmatrix}$.  $\ket{1}_\alpha$ and $\ket{0}_\beta$ represent the presence of a neutrino in flavor-mode $\alpha$ and its absence in flavor-mode $\beta$, respectively. The time evolution of, say an initial $\nu_e$ state, in a two-flavor mode system can be expressed as, 
	\begin{equation} \label{two_flavoreq}
		\ket{\nu_e(t)} = \widetilde{\mathcal{U}}_{ee}(t) \ket{10}_e + \widetilde{\mathcal{U}}_{e\mu}(t) \ket{01}_\mu.
	\end{equation} 
	The probability of detecting $\nu_e$ flavor state as a disappearance probability $P_d = \left| \widetilde{\mathcal{U}}_{ee} \right|^2$ and the probability of detecting $\nu_\mu$ flavor state as an appearance probability 
	$P_a = \left| \widetilde{\mathcal{U}}_{e\mu} \right|^2$.

Similarly, a three-flavor neutrino state can be mapped to a three-qubit state using the following occupation number representation of flavor states at $t=0$,
	\[
	\begin{aligned}
		\ket{\nu_e} &= \ket{1}_e \otimes \ket{0}_\mu \otimes \ket{0}_\tau \equiv \ket{100}_e ,\\
		\ket{\nu_\mu} &= \ket{0}_e \otimes \ket{1}_\mu \otimes \ket{0}_\tau \equiv \ket{010}_\mu ,\\
		\ket{\nu_\tau} &= \ket{0}_e \otimes \ket{0}_\mu \otimes \ket{1}_\tau \equiv \ket{001}_\tau .
	\end{aligned}
	\]
	Here,  $\ket{1}_e= \begin{pmatrix}
		0\\
		1
	\end{pmatrix}$ ,  $\ket{0}_\mu= \begin{pmatrix}
		1\\
		0
	\end{pmatrix}$, and  $\ket{0}_\tau= \begin{pmatrix}
		1\\
		0
	\end{pmatrix}$.
	
    The time evolution of a three-flavor neutrino state in the  computational basis, can be written as,
\begin{equation}
	\ket{\nu_\beta(t)} = \widetilde{\mathcal{U}}_{\beta e}(t)\ket{100}_e + \widetilde{\mathcal{U}}_{\beta \mu}(t)\ket{010}_\mu + \widetilde{\mathcal{U}}_{\beta \tau}(t)\ket{001}_\tau,
	\label{eq:three_flavor_state}
\end{equation}\\ 
where, $P_d = \left|\widetilde{\mathcal{U}}_{\beta\beta}\right|^2$ and $P_a = \sum_{\alpha\neq \beta}\left|\widetilde{\mathcal{U}}_{\beta \alpha}\right|^2$.
   
To quantify the degree of entanglement, we prepare a density matrix $\rho$, as this measure can be easily quantified using the appropriate density matrix. Since our study is based on the ongoing long-baseline experiments, NO$\nu$A and T2K, the initial neutrino beam is of muon flavor. Hence, for these long-baseline experiments, the time-evolved flavor state can be expressed by eq.~\ref{eq:three_flavor_state} with $\beta=\mu$.

As a result, the density matrix for the three-flavor neutrino system can be articulated as $\rho_\mu (t) = \ket{\nu_\mu (t)}\bra{\nu_\mu (t)}$, which can be linked with the matter-modified mixing matrix elements as follows, 
\begin{equation}
	\rho_\mu (T=t) = \begin{pmatrix}
0 & 0 & 0 & 0 & 0 & 0 & 0 & 0 \\
0 & ~~|\widetilde{\mathcal{U}}_{\mu\tau}|^2 & ~~\widetilde{\mathcal{U}}_{\mu\tau}\widetilde{\mathcal{U}}^*_{\mu\mu} & 0 & ~~\widetilde{\mathcal{U}}_{\mu\tau}\widetilde{\mathcal{U}}^*_{\mu e} & 0 & 0 & 0 \\
0 & ~~\widetilde{\mathcal{U}}_{\mu\mu}\widetilde{\mathcal{U}}^*_{\mu\tau} & ~~|\widetilde{\mathcal{U}}_{\mu\mu}|^2 & 0 & ~~\widetilde{\mathcal{U}}_{\mu\mu}\widetilde{\mathcal{U}}^*_{\mu e} & 0 & 0 & 0 \\
0 & 0 & 0 & 0 & 0 & 0 & 0 & 0 \\
0 & ~~\widetilde{\mathcal{U}}_{\mu e}\widetilde{\mathcal{U}}^*_{\mu\tau} & ~~\widetilde{\mathcal{U}}_{\mu e}\widetilde{\mathcal{U}}^*_{\mu\mu} & 0 & ~~|\widetilde{\mathcal{U}}_{\mu e}|^2 & 0 & 0 & 0 \\
0 & 0 & 0 & 0 & 0 & 0 & 0 & 0 \\
0 & 0 & 0 & 0 & 0 & 0 & 0 & 0 \\
0 & 0 & 0 & 0 & 0 & 0 & 0 & 0 \\
\end{pmatrix}.
	\label{eq:density_matrix}
\end{equation}

	\subsection{Total concurrence for neutrino flavor state}	
	The post measurement vectors for the two-flavor neutrino state defined in eq.~\ref{two_flavoreq} are ${\widetilde{\mathcal{U}}}_{ee}\ket{1}$ and ${\widetilde{\mathcal{U}}}_{e \mu} \ket{0}$, which form an orthogonal set of states. 
    Concurrence in this case can be defined as $ 2|{\widetilde{\mathcal{U}}}_{ee}||{\widetilde{\mathcal{U}}}_{e \mu}|$ and hence we get the squared concurrence as $4 P_a P_d$.

	For three-flavor oscillations, we have concurrence for the bipartitions $C_{\mu|e \tau}$, $C_{e|\mu \tau}$ and $C_{\tau| \mu e}$. The values of these bipartitions~\cite{nielsen_chuang_2010} for initial $\nu_\mu$ state can be given as follows, 
\begin{align}
C_{\mu|e\tau}
&=
\sqrt{
2\left[
1-
\left(|\widetilde{\mathcal{U}}_{\mu e}(t)|^2+|\widetilde{\mathcal{U}}_{\mu\tau}(t)|^2\right)^2
-
|\widetilde{\mathcal{U}}_{\mu\mu}(t)|^4
\right]
}
=
2|\widetilde{\mathcal{U}}_{\mu\mu}(t)|
\sqrt{1-|\widetilde{\mathcal{U}}_{\mu\mu}(t)|^2},
\nonumber\\[2mm]
C_{e|\mu\tau}
&=
\sqrt{
2\left[
1-
\left(|\widetilde{\mathcal{U}}_{\mu\mu}(t)|^2+|\widetilde{\mathcal{U}}_{\mu\tau}(t)|^2\right)^2
-
|\widetilde{\mathcal{U}}_{\mu e}(t)|^4
\right]
}
=
2|\widetilde{\mathcal{U}}_{\mu e}(t)|
\sqrt{1-|\widetilde{\mathcal{U}}_{\mu e}(t)|^2},
\nonumber\\[2mm]
C_{\tau|\mu e}
&=
\sqrt{
2\left[
1-
\left(|\widetilde{\mathcal{U}}_{\mu e}(t)|^2+|\widetilde{\mathcal{U}}_{\mu\mu}(t)|^2\right)^2
-
|\widetilde{\mathcal{U}}_{\mu\tau}(t)|^4
\right]
}
=
2|\widetilde{\mathcal{U}}_{\mu\tau}(t)|
\sqrt{1-|\widetilde{\mathcal{U}}_{\mu\tau}(t)|^2}.
\end{align}
   
As in the beam experiments, $\nu_\mu$ is mainly the initial flavor, we can project the general equation of total concurrence to the canvas of the three-flavor neutrino landscape. Using proper PMNS matrix elements, the modified definition of squared total concurrence can be expressed as\\
\begin{equation}
    C^{\mu^2} = 8 P_{\mu e}P_{\mu\mu} + 8 P_{\mu \tau}P_{\mu\mu} + 8 P_{\mu e}P_{\mu\tau}.
    \label{eq: Total_squared_concurrence_in_three_neutrino}
\end{equation}


 \section{Experimental setups and variation of total concurrence}
 \label{sec:ExperimentalDetails}

Long-baseline neutrino experiments play a central role in addressing some of the most fascinating questions in neutrino physics~\cite{Feldman:2012jdx, Agarwalla:2014fva, Rahaman:2022rfp, Diwan:2016gmz}. By sending intense and carefully tuned neutrino beams over distances of hundreds to thousands of kilometers, these experiments allow us to observe how neutrinos transform as they travel through matter. This setup provides a powerful means to explore key issues such as leptonic CP violation~\cite{Pascoli:2006ci, T2K:2019bcf, Rahaman:2022rfp, Agarwalla:2013avs, Ghosh:2013yon}, resolving the $\theta_{23}$ octant degeneracy~\cite{Agarwalla:2013ju, Agarwalla:2021bzs, Ghosh:2015ena}, establishment of the correct neutrino mass ordering~\cite{Dixit:2018kev, Petcov:2005rv, Agarwalla:2013qfa, Prakash:2013nra, Dutta:2012et}, and matter effects~\cite{Wolfenstein:1977ue, Mikheyev:1985zog, Mikheyev1986, Minakata:2015gra, Denton:2016wmg} in oscillations. Moreover, they offer an important avenue for probing possible physics beyond the Standard Model~\cite{Mena:2005ek, Feldman:2012jdx, Agarwalla:2014fva, Diwan:2016gmz, Farzan:2017xzy}. We commence our analysis with some comprehensive juxtaposition of the two well-established long-baseline experiments presently in operation, NO$\nu$A and T2K, highlighting their defining attributes and experimental sensitivities. First, we compute the total concurrence for neutrinos produced with an initial muon flavor using the NO$\nu$A and T2K experimental setups, and vary the oscillation parameters to obtain a common set of parameters for which the total concurrence is minimum for both experiments. Thereafter, with the aid of publicly available GLoBES~\cite{Huber:2004ka,Huber:2007ji} software, we vary selected parameters to fit NO$\nu$A and T2K data~\cite{Mikola:2024jnj,Wolcott:2024} while keeping other parameters to their values obtained from the total concurrence.

\subsection{An overview of the experimental setups of NO$\nu$A and T2K}
\label{setup-nova-t2k}
NuMI Off-axis $\nu_e$ Appearance (NO$\nu$A)~\cite{NOvA:2007rmc, NOvA:2021nfi}, is an ongoing USA-based long-baseline experiment with a baseline of 810 km, spanning from Fermilab to Ash River, Minnesota. This on-axis and narrow band experiment uses a 14 kt scintillator detector with nominal exposure of 58.8 kt$\cdot$MW$\cdot$yr and has been collecting data since 2014. The total 6-year run time is equally split between neutrino and antineutrino modes. The experiment typically runs with a beam power of about 700 kW, accumulating approximately $3.6 \times 10^{21}$ protons on target (P.O.T). Due to its $0.8^\circ$ off-axis orientation, the maximum flux is at 2 GeV, which is near its first oscillation maxima at 1.6 GeV. As signals, we consider the combined contributions of $\nu_\mu \rightarrow \nu_e$ together with $\bar{\nu}_\mu \rightarrow \bar{\nu}_e$ for the appearance channels, and $\nu_\mu \rightarrow \nu_\mu$ along with $\bar{\nu}_\mu \rightarrow \bar{\nu}_\mu$ for the disappearance channels. The systematic uncertainties of $5\%$ in signals and $10\%$ in backgrounds are considered for both the appearance and disappearance channels' events.

Tokai to Kamioka (T2K)~\cite{T2K:2023smv} is another ongoing narrow-band experiment in Japan, in which an intense $\nu_\mu$ beam produced at the J-PARC accelerator facility in Tokai traverses a distance of 295 km to reach the water Cherenkov detector Super-Kamiokande, with a fiducial volume of 22.5 kt. The experiment operates with a beam power of about 750 kW, corresponding to a total exposure of $7.8 \times 10^{21}$ protons on target, and has been collecting data since 2009. The total runtime of five years is equally divided between neutrino and antineutrino modes, leading to a nominal exposure of 84.4 kt$\cdot$MW$\cdot$yr. The $2.5^\circ$ off-axis configuration helps it to peak its flux at its first oscillation maxima at 0.6 GeV. The budget of systematic uncertainties for T2K is the same as mentioned for NO$\nu$A.

It is worth noting that the far detectors of both experiments are non-magnetized and therefore lack the capability to distinguish between neutrino and antineutrino interactions on an event-by-event basis. As a consequence, any wrong-sign contamination present in a given beam configuration, whether nominally neutrino or antineutrino, cannot be directly identified in the detector. Accordingly, in our analysis, we include the wrong-sign contributions as part of the signal. The average Earth matter density for both experiments is considered as 2.8 $gm/cm^3$.
Recently, NO$\nu$A and T2K presented their joint analysis~\cite{T2K:2025wet}, and in our synergistic study of these two experiments, we employ the benchmark oscillation parameter values provided by the collaborations.
\begin{table}[htb!]
\centering
	\caption{The benchmark values of the oscillation parameters and their corresponding 3$\sigma$ ranges  following the ref.~\cite{Esteban:2024eli}.}
\resizebox{\columnwidth}{!}{%
\begin{tabular}{|c|c|c|c|c|c|c|}
\hline \hline
\multirow{2}{*}{\textbf{Parameter}} & $\Delta m^2_{21}/10^{-5}$ & $\theta_{12}$ & $\theta_{13}$ & $\theta_{23}$ & $\Delta m^2_{31}/10^{-3}$  & $\delta_{\text{CP}}$ \\
				&($\mathrm{eV^{2}}$) & $(^\circ)$ & $(^\circ)$ & &($\mathrm{eV^{2}}$)& ($^\circ$)\\
				\hline \hline
				\multirow{2}{*}{\textbf{NMO}} & \boldmath$7.49$ & \boldmath$33.68$ & \boldmath$8.56$ & \boldmath$43.3$ & \boldmath$2.513$ & \boldmath$212$\\
				\cline{2-7}
				& 6.92 - 8.05 & 31.63 - 35.95 & 8.19 - 8.89 & 41.3 - 49.9 & 2.451 - 2.578 & 124 - 364\\
				\hline \hline
                \multirow{2}{*}{\textbf{IMO}} & \boldmath$7.49$ & \boldmath$33.68$ & \boldmath$8.59$ & \boldmath$47.9$ & \boldmath$-2.484$ & \boldmath$274$\\
				\cline{2-7}
				& 6.92 - 8.05 & 31.63 - 35.95 & 8.25 - 8.93 & 41.5 - 49.8 & (-2.547)- (-2.421) & 201 - 335\\
				\hline \hline
			\end{tabular}
			}
			\label{table:NuFit}
	\end{table}


\subsection{Common benchmark oscillation parameter sets between NO$\nu$A and T2K }
NO$\nu$A and T2K are two forefront long-baseline experiments that operate in different energy regimes and probe distinct first oscillation maxima. Being off-axis experiments, their fluxes peak at different energies due to differences in their respective off-axis angles relative to the beam pipes. Consequently, their sensitivities to total concurrence can differ, reflecting the underlying differences in their energy coverage.
To scan the effect of entanglement measure on the sensitivity to oscillation parameters, we need to focus on the dependence of the total concurrence directly on the oscillation parameters, however, being minimally affected by the difference of their operational energy ranges. To achieve this, we calculate the total concurrence within the energy range corresponding to the full width at half maxima (FWHM) around their  
peak flux, varying the oscillation parameters within their $3\sigma$ uncertainties reported in~\cite{Esteban:2024eli} and listed in table~\ref{table:NuFit}.
Further, among these values, we selected a common set of six neutrino oscillation parameters that yielded the minimum total concurrence achieved for the T2K and NO$\nu$A experiments. These parameter values are detailed in table~\ref{tab:NMO_concurrence_combined} and table~\ref{tab:IMO_concurrence_combined}, in case of normal mass ordering (NMO) and inverted mass ordering (IMO), respectively. The minimum value of total concurrence, particularly its local minima, is significant for the clear extraction of oscillation parameters. This is because these minima correspond to the energy window where the neutrino flavor state is closest to being separable. In the context of another long-baseline neutrino experiment under construction, known as Deep Underground Neutrino Experiment (DUNE)~\cite{DUNE:2020jqi}, it has been reported that this local minima of total concurrence indicate a nearly separable (though not fully separable) neutrino flavor state~\cite{Banerjee:2024lih}. This characteristic makes it especially suitable for the precise extraction of oscillation parameters. 

We found two common oscillation parameter sets (set-1 and set-2) corresponding to the concurrence local minima for T2K and NO$\nu$A in the case of NMO, while only one such set in case of IMO. These parameters serve as our benchmark choices, significantly reducing the impact of the discrepancies between the operational energy windows of NO$\nu$A and T2K.
Such a prescription of choosing this common set of oscillation parameters also ensures that our results are less biased toward a particular global-fit preference.\\
\begin{table}[htbp]
\centering
\caption{Common sets of oscillation parameters between T2K and NO$\nu$A corresponding to minimum total concurrence within the allowed $3\sigma$ ranges of the global fit~\cite{Esteban:2024eli}, and within their respective energy ranges containing FWHM of flux, assuming NMO.}

\resizebox{\textwidth}{!}{%
\begin{tabular}{|c|c|c|c|c|c|c|c|c|c|}
\hline\hline
\textbf{Type} & \textbf{Set} & \textbf{Experiment} & $\Delta m^2_{21}/(10^{-5})$ & $\theta_{12}$ & $\theta_{13}$ & $\theta_{23}$ & $\Delta m^2_{31}/(10^{-3})$ & $\delta_{CP}$ & \textbf{Total concurrence} \\
\textbf{of extrema} & & \textbf{[E (GeV)]} & \textbf{[eV$^2$]} & & &  & \textbf{[eV$^2$]} & & \textbf{(NMO)}\\
\hline

\multirow{4}{*}{\textbf{Min}} 
& Set-1 & NO$\nu$A [1.73] 
& \multirow{2}{*}{8.05} 
& \multirow{2}{*}{35.9} 
& \multirow{2}{*}{8.19} 
& \multirow{2}{*}{45.0} 
& \multirow{2}{*}{2.574} 
& \multirow{2}{*}{88} 
& 0.165057 \\
\cline{10-10}
&       & T2K [0.62]     
& & & & & & & 0.131989 \\
\cline{2-10}

& Set-2 & NO$\nu$A [2.01] 
& \multirow{2}{*}{6.93} 
& \multirow{2}{*}{35.9} 
& \multirow{2}{*}{8.19} 
& \multirow{2}{*}{45.0} 
& \multirow{2}{*}{2.574} 
& \multirow{2}{*}{92} 
& 0.40973 \\
\cline{10-10}
&       & T2K [0.685, 0.75] 
& & & & & & & 0.260415 (0.447701) \\

\hline

\hline\hline
\end{tabular}
}
\label{tab:NMO_concurrence_combined}
\end{table}    

\begin{table}[htbp]
\centering
\caption{Common oscillation parameters between T2K and NO$\nu$A corresponding to minimum total concurrence within the allowed $3\sigma$ ranges of the global fit~\cite{Esteban:2024eli}, and within their respective energy ranges containing FWHM of flux, assuming IMO.}

\resizebox{\textwidth}{!}{%
\begin{tabular}{|c|c|c|c|c|c|c|c|c|}
\hline\hline
\textbf{Type} & \textbf{Experiment} & $\Delta m^2_{21}/(10^{-5})$ & $\theta_{12}$ & $\theta_{13}$ & $\theta_{23}$ & $\Delta m^2_{31}/(10^{-3})$ & $\delta_{CP}$ & \textbf{Total concurrence} \\
\textbf{of extrema} & \textbf{[E (GeV)]} & \textbf{[eV$^2$]} & & & & \textbf{[eV$^2$]} & & \textbf{(IMO)} \\
\hline

\multirow{2}{*}{\textbf{Min}}
& NO$\nu$A [1.93--2.03] 
& \multirow{2}{*}{8.05}
& \multirow{2}{*}{35.9}
& \multirow{2}{*}{8.25}
& \multirow{2}{*}{45.0}
& \multirow{2}{*}{-2.547}
& \multirow{2}{*}{88}
& 0.216--0.319 \\
\cline{9-9}
& T2K [0.62] 
& & & & & & & 0.0998 \\

\hline

\hline\hline
\end{tabular}
}
\label{tab:IMO_concurrence_combined}
\end{table}

\begin{figure}[H]
	\centering 
    \includegraphics[width=0.95\linewidth]{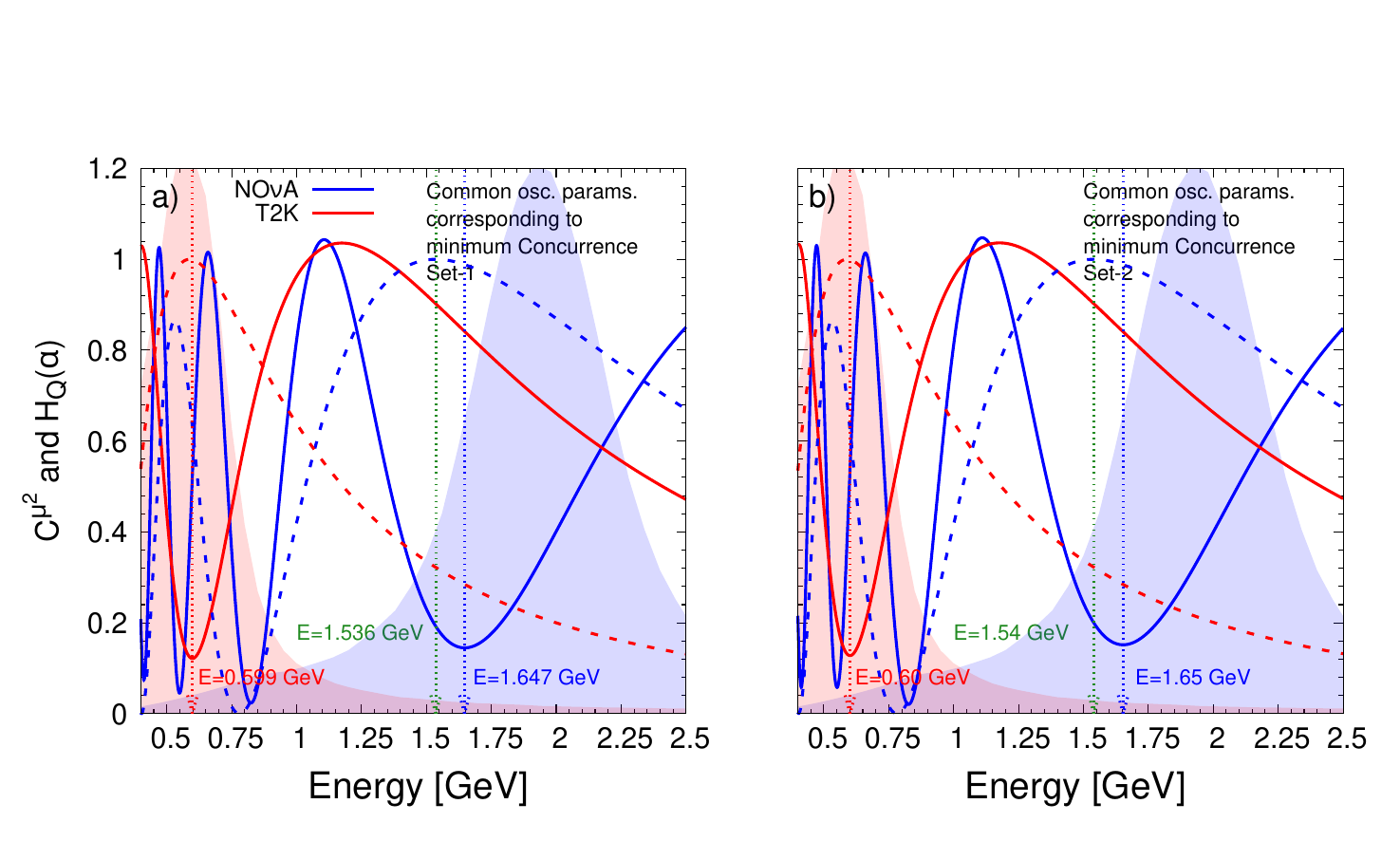}\\ 
	\includegraphics[width=0.95\linewidth]{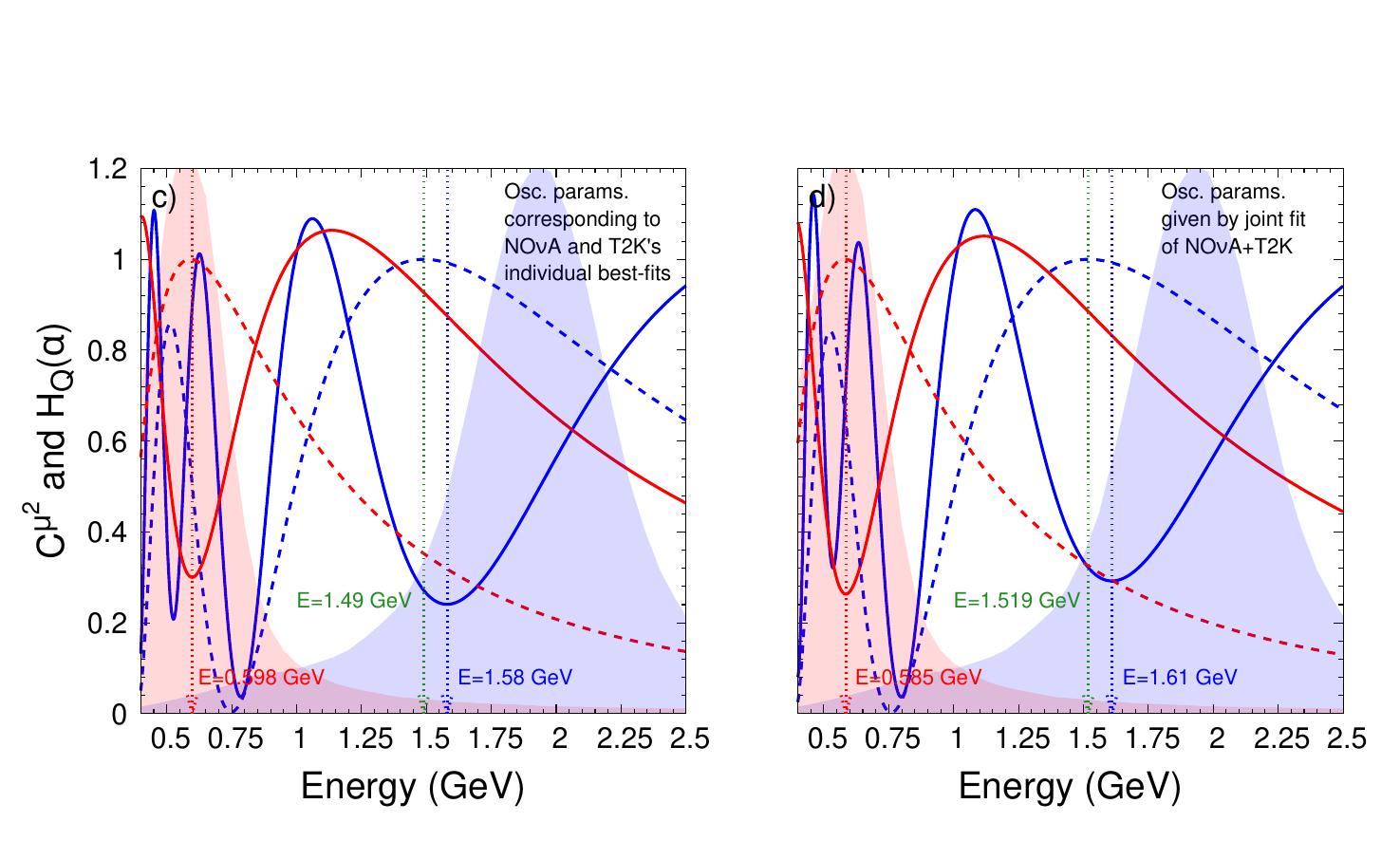}
  
\caption{(a) [(b)] The common oscillation parameters corresponding to minimum concurrence, set-1 [set-2] of Table~\ref{tab:NMO_concurrence_combined}, 
(c) the benchmark choices given by NO$\nu$A and T2K, individually~\cite{T2K:2025wet}, and (d) the benchmark values given by the joint analysis of NO$\nu$A and T2K~\cite{T2K:2025wet}, assuming NMO. The solid lines depict total concurrence ($C^{\mu^2}$) scaled by a factor of 1/2, whereas the dashed lines represent the normalized Quantum Fisher Information (QFI) ($H_Q(\alpha)$ for $\alpha=\delta_{CP}$)~\cite{Nogueira:2016qsk} for T2K (red) and NO$\nu$A (blue) as a function of neutrino energy. The red (blue) vertical dotted lines indicate the neutrino energies corresponding to the local minimum of the total concurrence for T2K (NO$\nu$A), while the olive dotted lines denote the energies at which the QFI is maximum. For T2K, the concurrence minimum and QFI maximum coincide, resulting in a single vertical line.} 
	\label{fig:2_conc_vs_energy}
\end{figure}
Figure~\ref{fig:2_conc_vs_energy}  portrays the behavior of the total concurrence as a function of neutrino energy for NO$\nu$A and T2K. For the upper panel of figure~\ref{fig:2_conc_vs_energy}, the benchmark oscillation parameters are taken from table~\ref{tab:NMO_concurrence_combined}. The left column corresponds to set-1, while the right column corresponds to set-2, both for the case of NMO. We also present the variation in total concurrence with energy for the best-fit oscillation parameter values reported in~\cite{T2K:2025wet} by the T2K and NO$\nu$A collaborations, individually (left) and from their combined fit (right) in the lower panel of figure~\ref{fig:2_conc_vs_energy}.  
The blue and red solid lines represent total concurrence ($C^{\mu^2}$), given in eq.~\ref{eq: Total_squared_concurrence_in_three_neutrino}, for the NO$\nu$A and T2K experiments, respectively. In each panel, the blue (NO$\nu$A) and red (T2K) shaded regions illustrate the corresponding experimental flux measurements.

In the upper panels of figure~\ref{fig:2_conc_vs_energy}, we observe that the concurrence local minima for T2K occur at nearly the same energy level, approximately 0.599 GeV for set-1 (left panel) and around 0.60 GeV for set-2 (right panel). In contrast, the concurrence local minima for NO$\nu$A are shifted to energy values that align with its FWHM region of the maximum neutrino flux, specifically around 1.647 GeV for set-1 and 1.65 GeV for set-2. These energy values are closer to the peak flux energy for NO$\nu$A at 2 GeV compared to the local concurrence minima observed in the lower panel, which are at approximately 1.58 GeV and 1.61 GeV. As a result, selecting a set of common oscillation parameters for T2K and NO$\nu$A based on the minimum concurrence achieved by both experiments effectively shifts the local minima of total concurrence for each experiment to an energy range within FWHM of their respective fluxes, where most of the neutrino events are detected, thus increasing statistical significance. We checked that a similar conclusion can be drawn for IMO by adopting the common set of oscillation parameters for the NO$\nu$A and T2K configurations, as listed in table~\ref{tab:IMO_concurrence_combined}. 
 
We refer to this strategy of total concurrence local-minima-shift (LMS) within the FWHM regions for each experiment, the LMS scheme. This scheme enhances the likelihood of accurately extracting the oscillation parameters. In the next section, we use the benchmark values of oscillation parameters given in table~\ref{tab:NMO_concurrence_combined} (for NMO) and table~\ref{tab:IMO_concurrence_combined} (for IMO) to present our results regarding the oscillation parameter estimation. 

\subsection{Relationship between total concurrence and quantum Fisher information}
To justify why the local minima of total concurrence ${C^\mu}^2$ help us extract neutrino oscillation parameters more precisely, we need to explore the connection among three core concepts: flavor entanglement (defined here by $C^{\mu^2}$), parameter sensitivity of state, and standard experimental data analysis ($\chi^2$). In the previous subsection, we discussed the significance of local minima of total concurrence for the accurate extraction of oscillation parameters. These local minima represent nearly separable (but not fully) neutrino flavor states, which aid in the extraction of oscillation parameters.  To quantify the parameter-sensitivity aspect of this connection, we describe neutrino oscillations within the framework of quantum estimation theory (QET), closely related to quantum metrology, \cite{1969JSP.....1..231H, holevo2011probabilistic, Paris:2008zgg, Albarelli:2020pec}, where the evolving flavor state $\ket{\nu_{\alpha}(t)}$ acts as a carrier of information about the underlying oscillation parameters through its dependence on the PMNS mixing matrix. The ultimate precision with which these parameters can be estimated is characterized by the quantum Fisher information (QFI), which quantifies the maximum information that can, in principle, be extracted from the quantum state. The associated Cramér–Rao bounds then set the minimum attainable uncertainty in parameter estimation, with the QFI providing the ultimate quantum precision limit. For a neutrino state $\rho (\alpha)$ depending on an oscillation parameter $\alpha$, the QFI quantifier, $H_Q(\alpha)$ determines the ultimate precision allowed by a quantum mechanical state through the Cramér–Rao bound,
\begin{equation}
    \Delta \alpha \geq \frac{1}{\sqrt{N H_Q(\alpha)}}
    \label{eq:QFI_bound},
\end{equation}
with N denoting the number of independent measurements (event counts for neutrinos) and $H_{Q}(\alpha)$ is expressed as, 
\begin{equation}
    H_{Q}(\alpha) = 4\left(
\langle \partial_\alpha \psi_\alpha | \partial_\alpha \psi_\alpha \rangle
-
\left|\langle \psi_\alpha | \partial_\alpha \psi_\alpha \rangle\right|^2
\right).
\label{eq:QFI}
\end{equation} 
\noindent
Thus, we will examine the following two key points to determine whether there is any connection between them:
\begin{enumerate}
    \item \textit{Local minima of entanglement measure ${C^\mu}^2$} showing exactly where the three neutrino flavor modes are closest to being unentangled (separable). 
    \item \textit{Highest parameter sensitivity corresponding to $H_Q(\alpha)$ Maxima} where the wave function is highly dynamic and undergoes its sharpest physical variations. 
    \item \textit{Smaller error bars ($\chi^2$ contours)} based on Eq.~\eqref{eq:QFI_bound}, i.e., a higher $H_Q(\alpha)$ mathematically forcing the lower uncertainty $\Delta\alpha$ in the parameter. 
\end{enumerate}
For this purpose, in figure~\ref{fig:2_conc_vs_energy}, we have plotted $H_{Q}(\alpha)$ for $\alpha=\delta_{CP}$ as dashed-blue and dashed-red curves for NO$\nu$A and T2K, respectively. We observed a relationship between the local minima of $C^{\mu^2}$, as defined in eq. \ref{eq: Total_squared_concurrence_in_three_neutrino}, and the maxima of $H_{Q}(\alpha)$, noting that they coincide closely with each other. A similar connection has been reported between the von Neumann entropy-based entanglement measure and the \textit{classical} Fisher Information associated with flavor measurements in~\cite{Nogueira:2016qsk} for the two-flavor neutrino oscillation scenario. However, in their work, the QFI is found to be constant and no such relation was observed between local minima of the entanglement measure and the maxima of QFI. In this work, we demonstrated the relationship between local minima of the tripartite entanglement measure $C^{\mu^2}$ and the maxima of QFI quantifier $H_Q(\alpha)$ for the first time in the context of three-flavor neutrino oscillations. Regions of maximum QFI indicate areas where small changes in the oscillation parameter result in more distinguishable quantum states. This, in turn, leads to a smaller achievable uncertainty. Hence, in the context of neutrino oscillations, QFI quantifies how sensitive the state, undergoing flavor oscillations, is to intrinsic parameters. This observation of the close alignment between the maxima of $H_{Q}(\alpha)$ and the local minima of $C^{\mu^2}$ strongly supports the notion that local minima of total concurrence aid in effective parameter extraction.
 
In figure~\ref{fig:2_conc_vs_energy}, it can be observed that the maxima of $H_{Q}(\alpha)$ for NO$\nu$A (at $E=1.49$ GeV for the benchmarks from NuFIT 6.0) also shifts, landing at $E=1.536$ GeV for set-1 and at $E=1.54$ GeV for set-2 in table~\ref{tab:NMO_concurrence_combined} under our LMS scheme. This movement occurs alongside the shift of the local minima of $C^{\mu^2}$, bringing them both closer to the energy value of the peak flux for NO$\nu$A. In contrast, T2K exhibits no significant shift, as its peak flux already aligns very closely with the local minima of $C^{\mu^2}$ and the maxima of $H_{Q}(\alpha)$. 
This alignment suggests that by implementing our LMS scheme, we can synchronize the peak sensitivity (the maximum of $H_Q$) with the FWHM region of the experiment, where the event count $N$ is highest. By maximizing both $N$ and $H_Q$ simultaneously, we minimize the overall uncertainty as expressed in Eq.~\eqref{eq:QFI_bound}. This indicates reduced allowed $1\sigma$ boundaries and production of tighter $\chi^2$ confidence contours, as presented in our results section.

It can be seen that the alignment between $C^{\mu^2}$ local-minima and $H_Q(\alpha)$ maxima is a bit off for NO$\nu$A, which occurs due to the profound matter effect for this experiment. However, the LMS scheme shifts both $C^{\mu^2}$ local-minima and the $H_Q(\alpha)$ maxima closer to the peak flux energy. 

Recent studies have further emphasized the usefulness of QET for assessing the intrinsic and experimentally accessible information on neutrino oscillation parameters~\cite{Ignoti:2025rxr,Frugiuele:2026yeq,Chundawat:2026lcm,Huang:2026bws,Yadav:2026mnw,Lu:2026wzr}. In particular, comparisons between the QFI and the Fisher information associated with flavor measurements indicate that the present uncertainty on $\delta_{\rm CP}$ is not solely dictated by the information encoded in the neutrino quantum state, but also by how efficiently this information is accessed experimentally, with the extraction efficiency depending strongly on the oscillation region~\cite{Ignoti:2025rxr,Frugiuele:2026yeq,Chundawat:2026lcm}. More general QFI-matrix analyses have further highlighted the role of correlations among oscillation parameters and the associated degeneracy structure in determining their ultimate precision bounds~\cite{Huang:2026bws,Yadav:2026mnw}. These observations complement our analysis, where the close correspondence between concurrence minima and QFI maxima identifies intrinsically information-rich regions, while the LMS scheme aims to bring such regions closer to the high-statistics part of the experimental spectrum.

\subsection{Construction of the $\chi^2$ }
The sensitivity is quantified using the Poisson likelihood-based $\chi^2$ function~\cite{Baker:1983tu}, from which we extract the median sensitivity~\cite{Cowan:2010js} within a frequentist framework~\cite{Blennow:2013oma}. We define our $\chi^2$ in the following fashion,
\begin{equation}
\chi^2 (\vec{\omega}, \kappa_{s}, \kappa_{b,l}) =
\underset{(\vec{\lambda}, \kappa_{s}, \kappa_{b,l})}{\mathrm{min}}
\left[
2\sum_{i=1}^{n}
\left(
\tilde{y}_i - x_i - x_i \ln \frac{\tilde{y}_i}{x_i}
\right)
+ \kappa_s^2 + \sum_{l}\kappa_{b,l}^2
\right],
\label{eq:chi2}
\end{equation}
where the summation extends over $n$ reconstructed energy bins. The minimization is carried out over the nuisance parameters ($\kappa_s$ and $\kappa_{b,l}$) and the set $\vec{\lambda}$ of oscillation parameters included in the marginalization; the specific choice of $\vec{\lambda}$ will be specified in the relevant subsections.
The expected number of events in the $i$-th bin is written as,
\begin{equation}
\tilde{y}_i(\vec{\omega}, \kappa_{s}, \kappa_{b,l}) =
N^{\mathrm{th}}_i(\vec{\omega}) \left(1 + \pi^s \kappa_s \right)
+ \sum_{l} N^b_{i,l}(\vec{\omega}) \left(1 + \pi^b \kappa_{b,l} \right),
\label{eq:chi}
\end{equation}
where $N^{\mathrm{th}}_i(\vec{\omega})$ denotes the predicted signal yield for oscillation parameters
 $\vec{\omega} = \{\theta_{23}, \theta_{13}, \theta_{12},\\ \Delta m^2_{21}, \Delta m^2_{31}, \delta_{\mathrm{CP}}\}$.
The term $N^b_{i,l}(\vec{\omega})$ represents the contribution from the $l$-th background component. While neutral-current backgrounds are independent of the oscillation parameters, charged-current backgrounds inherit a dependence on $\vec{\omega}$. The coefficients $\pi^s$ and $\pi^b$ encode the relative systematic uncertainties on the signal and background components, respectively, with $\kappa_s$ and $\kappa_{b,l}$ being the corresponding pull parameters. All systematics are treated as uncorrelated, and identical values are assumed for neutrino and antineutrino running modes. 

The binned data entering the likelihood are constructed as $x_i = N^{\mathrm{ex}}_i + N^b_{i,l}$, where $N^{\mathrm{ex}}_i$ denotes the observed charged-current signal events in the $i$-th bin. These include contributions from both disappearance channels ($\nu_\mu \to \nu_\mu$, $\bar{\nu}_\mu \to \bar{\nu}_\mu$) and appearance channels ($\nu_\mu \to \nu_e$, $\bar{\nu}_\mu \to \bar{\nu}_e$), while $N^b_{i,l}$ accounts for the associated background contributions.
Our numerical analysis is performed using the \texttt{GLoBES} framework~\cite{Huber:2002mx, Fogli:2002pt, Gonzalez-Garcia:2004pka}. In generating the pseudo-data, we assume NMO, consistent with the current global analyses that exhibit a preference for NMO at the level of about $2.5\sigma$~\cite{Esteban:2024eli}. During the fit, all relevant oscillation parameters are varied within their $3\sigma$ ranges, as summarized in table~\ref{table:NuFit}.

\section{Results}
\label{sec:Results}
We present our results in the context of oscillation parameter estimation, with a primary focus on the $CP$-phase $\delta_{CP}$, the mixing angle $\theta_{23}$, and the mass-squared difference $\Delta m_{31}^2$. Additionally, we provide results showing how our LMS scheme affects the sensitivity of both experiments to detecting $CP$ violation, resolving the $\theta_{23}$ octant, and determining the mass ordering. Our analysis considers the total contributions from neutrino and antineutrino events.

\subsection{Influence of LMS scheme on allowed region in ($\sin^2\theta_{23}-\delta_\mathrm{CP}$) plane} 

\begin{figure}[htb!]
	\centering 
 \includegraphics[width=\linewidth]{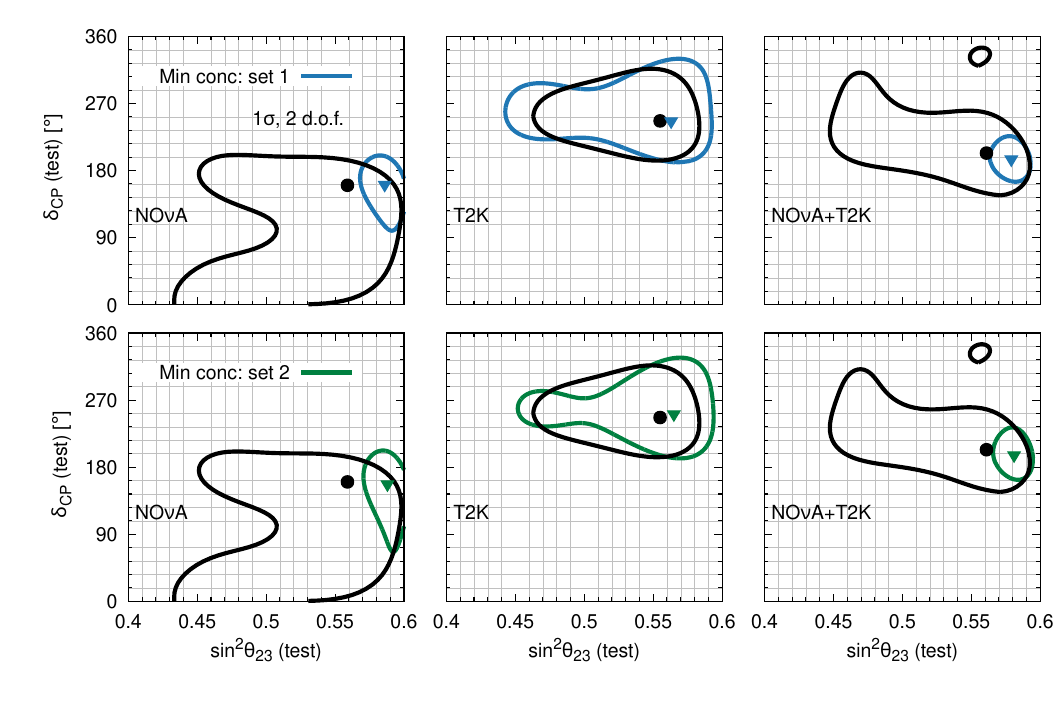}
	\caption{\footnotesize{These plots show the bounds in ($\sin^2\theta_{23}-\delta_\mathrm{CP}$) plane given by NO$\nu$A (left panel), T2K (middle panel), and NO$\nu$A+T2K (right panel). The contours are at $1\sigma$ C.L. for 2 d.o.f. Assuming a true NMO, the blue (upper row) and green (lower row) lines depict contours where the remaining oscillation parameters are fixed to the shared T2K and NO$\nu$A test values listed in table~\ref{tab:NMO_concurrence_combined} for set-1 and set-2, respectively. The lower triangular solid points represent the values of $(\sin^2\theta_{23},\delta_{\mathrm{CP}})$ for which $\Delta\chi^2$ is zero. The solid black lines are $1\sigma$ C.L. contours taken from the recent NO$\nu$A data~\cite{NOvA:2021nfi} (for left panel), recent T2K data~\cite{T2K:2023smv} (for middle panel) and recent joint result of T2K and NO$\nu$A~\cite{T2K:2025wet} (for right panel) respectively. The black solid dots represent the best-fit values as given by the collaborations~\cite{T2K:2025wet}. 
    } }
	\label{fig:Contour_min_concurrence_NMO}
\end{figure}

In this section, we investigate the effect of LMS scheme on the phenomenological perspective of possible correlations between $\sin^2\theta_{23}$ and $\delta_\mathrm{CP}$ in NO$\nu$A and T2K. To ensure the same footing of NO$\nu$A and T2K (as discussed earlier), our hypothesis is kept fixed at the set of common oscillation parameters corresponding to the local minima of total concurrence [see table~\ref{tab:NMO_concurrence_combined}]. We utilize the real data obtained by the collaborations as true data points. In figure~\ref{fig:Contour_min_concurrence_NMO}, we scan the parameter space ($\sin^2\theta_{23}-\delta_\mathrm{CP}$) as these are the two most uncertain parameters in the neutrino oscillation sector. This analysis includes results for NO$\nu$A (left panel), T2K (middle panel), and a combined fit for NO$\nu$A+T2K (right panel). To obtain this figure, our $\Delta\chi^2$ is defined as,
\begin{equation}
 \begin{aligned}
 \Delta \chi^2 = \underset{(\kappa_s,\,\kappa_b,\,\Delta m^2_{31})}{\mathrm{min}} \bigg[
    \chi^2\big(\sin^2\theta_{23}^{\mathrm{test}} \in [0.4:0.6],\, \delta_{\mathrm{CP}}^{\mathrm{test}} \in [0^\circ:\,360^\circ]\big) \\
    {}- \chi^2\big(\sin^2\theta_{23}^{\mathrm{true}},\,\delta_{\mathrm{CP}}^{\mathrm{true}} \in \mathrm{Real\,\, Data}\big)\bigg],
\end{aligned}
\label{eq:chi2-sensitivity-contour-real_th23_dcp}
\end{equation}
where we marginalize over systematic uncertainties of signals and backgrounds, as mentioned in Section~\ref{setup-nova-t2k}, and the $3\sigma$ range of $\Delta m^2_{31}$ as mentioned in table~\ref{table:NuFit}.  
In figure~\ref{fig:Contour_min_concurrence_NMO}, we project the  $\Delta\chi^2$ given in eq.~\ref{eq:chi2-sensitivity-contour-real_th23_dcp} in the parameter space at the $1\sigma$ C.L. for the parameter values (except $\theta_{23}$ and $\delta_{CP}$) given in table~\ref{tab:NMO_concurrence_combined} corresponding to our LMS scheme for NMO, in solid blue (for set-1) and green (for set-2) lines.
The blue and green triangular dots represent the coordinates in the above-mentioned parameter space for which $\Delta\chi^2=0$. The black solid lines and the circular black dots represent the favored regions at $1\sigma$ C.L. and the best-fit values, as reported by the recent NO$\nu$A and T2K collaborations~\cite{T2K:2025wet}. For NO$\nu$A, it can be seen that the 1$\sigma$ allowed region in the ($\sin^2 \theta_{23}-\delta_{CP}$) parameter space given in colored (blue and green) solid lines is significantly reduced compared to the region enclosed with black solid lines representing the NO$\nu$A collaboration observation at 1$\sigma$ C.L. In the case of T2K, there is a marginal difference in the regions enclosed by the colored and black lines. This can be explained by the fact that, within our LMS scheme, the local minima of concurrence is significantly shifted closer to the energy corresponding to the maximum flux for NO$\nu$A. In contrast, for T2K, the observed shift is less significant, as the T2K setup already places the peak flux quite closer to the local minima of concurrence. Furthermore, when considering the combined fit of NO$\nu$A and T2K data, we observe additional improvements due to the increased amount of statistical data in this scenario. Moreover, the best-fit points, denoted by colored triangular dots, are in very good agreement with the observed best-fit values from NO$\nu$A and T2K individually and in the combined fit for NO$\nu$A+T2K. Interestingly, the oscillation parameters obtained from the minimum concurrence (both set-1 and set-2 of table~\ref{tab:NMO_concurrence_combined}) resolve the $\theta_{23}$ octant degeneracy and show a preference for the higher octant in the left and right columns of figure~\ref{fig:Contour_min_concurrence_NMO}. In contrast, the middle panel, corresponding to the T2K experimental configuration, does not resolve the $\theta_{23}$ octant degeneracy, exhibiting nearly equal sensitivity to both the lower and higher octants. We find that, in the presence of minimum total concurrence corresponding to the benchmarks of set-1, the co-ordinates of the best-fit in ($\sin^2\theta_{23}-\delta_\mathrm{CP}$) plane are ($0.579^{+0.0139}_{-0.0154},\,{194^{+32}_{-29}}\,^\circ$), and for set-2, they are ($0.581^{+0.0136}_{-0.0150},\,{195^{+38}_{-32}}\,^\circ$); whereas the joint collaboration of NO$\nu$A and T2K~\cite{T2K:2025wet} gives their best-fit at ($0.561^{+0.021}_{-0.039},\,{203^{+63}_{-38}}\,^\circ$).

\begin{figure}[htb!]
	\centering 
 \includegraphics[width=\linewidth]{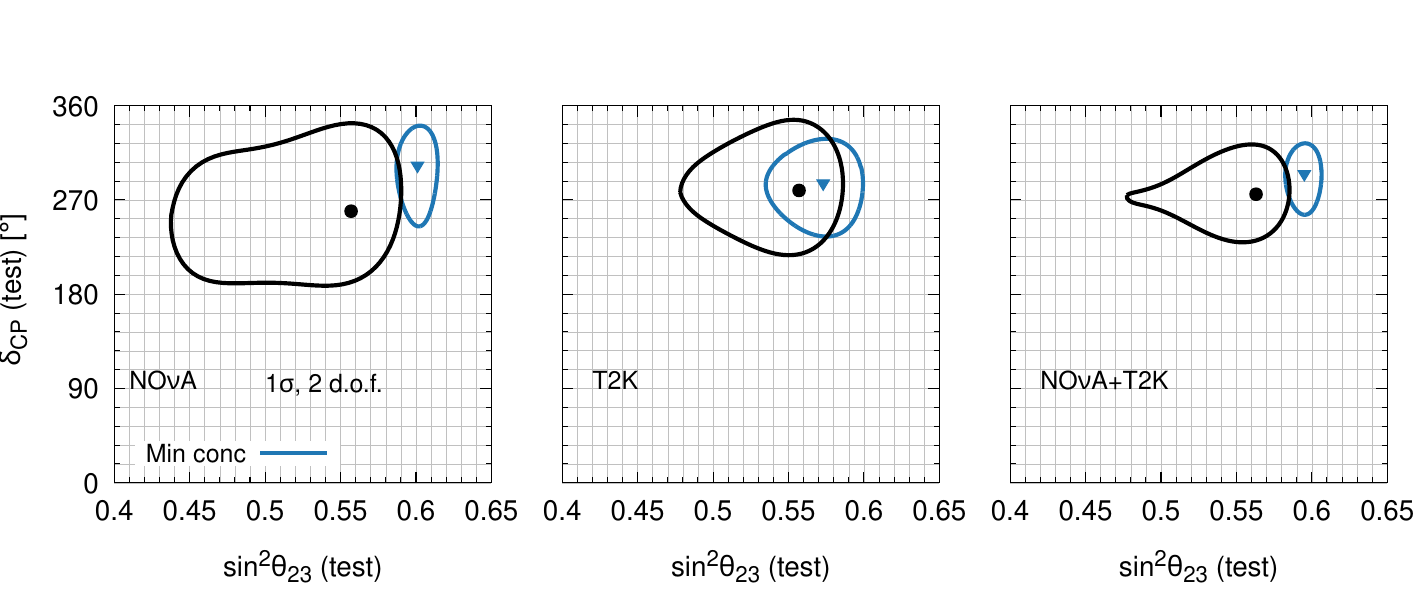}
	\caption{\footnotesize{Same as figure~\ref{fig:Contour_min_concurrence_NMO} using the common oscillation parameter set given in table~\ref{tab:IMO_concurrence_combined} for true IMO. 
    }} 
	\label{fig:Contour_1sigma_IMO_dcp_th23_conc}
\end{figure}

We also conduct the same analysis for the true IMO using the oscillation parameters listed in table~\ref{tab:IMO_concurrence_combined} and presented the results in figure~\ref{fig:Contour_1sigma_IMO_dcp_th23_conc}. This figure contains only three panels, as we found only one set of common oscillation parameters under our LMS scheme for this scenario. A similar observation and the explanation as mentioned for figure~\ref{fig:Contour_min_concurrence_NMO} fits for figure~\ref{fig:Contour_1sigma_IMO_dcp_th23_conc}. In figure~\ref{fig:Contour_1sigma_IMO_dcp_th23_conc}, the results obtained from our analysis consistently favor the higher octant across all panels, thereby resolving the $\theta_{23}$ octant degeneracy. We find that, in the presence of minimum total concurrence corresponding to the benchmarks of table~\ref{tab:IMO_concurrence_combined}, the co-ordinates of the best-fit in ($\sin^2\theta_{23}-\delta_\mathrm{CP}$) plane are ($0.595^{+0.0115}_{-0.0126},\,{294^{+30}_{-38}}\,^\circ$) for IMO; whereas the joint collaboration of NO$\nu$A and T2K~\cite{T2K:2025wet} gives their best-fit at ($0.563^{+0.021}_{-0.039},\,{275^{+31}_{-27}}\,^\circ$).

In figure~\ref{fig:Contour_min_concurrence_NMO}, the $\Delta\chi^2$ values, at the best-fit points, obtained under NMO hypothesis are 64.91, 94.90, and 163.92 for NO$\nu$A, T2K, and the combined NO$\nu$A+T2K analysis, respectively. In comparison, the corresponding values in case of IMO, shown in figure~\ref{fig:Contour_1sigma_IMO_dcp_th23_conc}, are 82.45, 102.291, and 186.45. Since the NMO hypothesis yields consistently lower $\Delta\chi^2$ values, our results indicate a preference for NMO over IMO.

\subsection{Influence of LMS scheme on allowed region in ($\sin^2\theta_{23}-\Delta m^2_{31}$) plane}

\begin{figure}[htb!]
	\centering 
 \includegraphics[width=\linewidth]{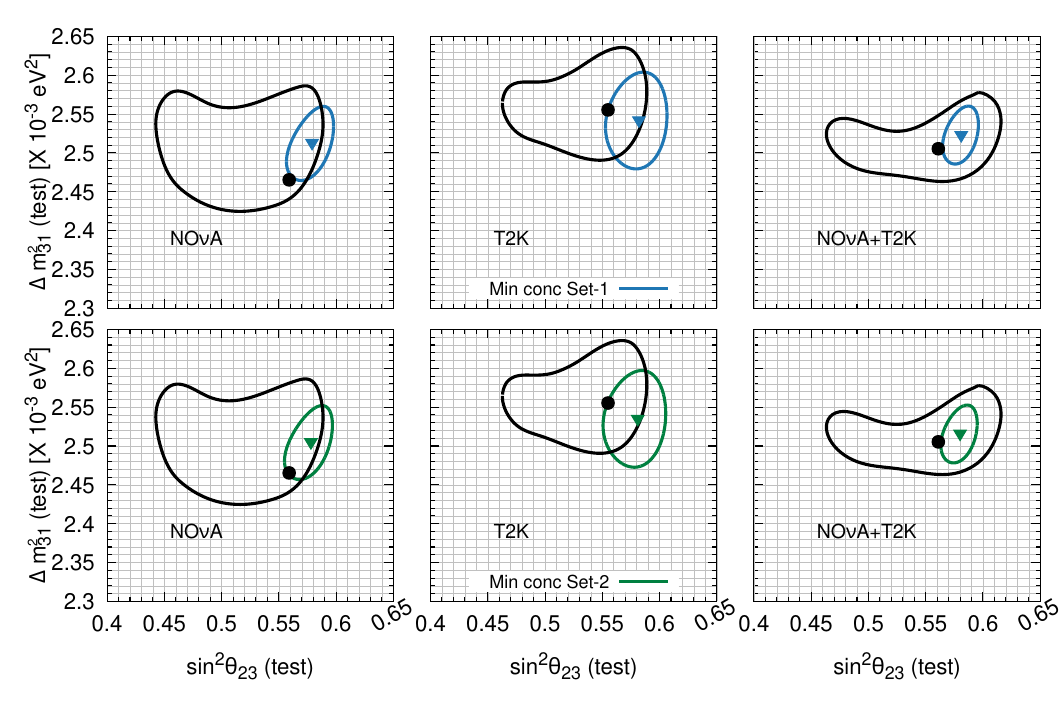}
	\caption{\footnotesize{These plots show the bounds in ($\sin^2\theta_{23}-\Delta m^2_{31}$) plane given by NO$\nu$A (left panel), T2K (middle panel), and NO$\nu$A+T2K (right panel) at $1\sigma$ C.L. for 2 d.o.f. Assuming a true NMO, the blue (upper row) and green (lower row) lines depict contours where the remaining oscillation parameters are fixed to the shared T2K and NO$\nu$A test values listed in table~\ref{tab:NMO_concurrence_combined} for set-1 and set-2, respectively.
The lower triangular solid points represent the values of $(\sin^2\theta_{23},\Delta m^2_{31})$ for which the value of $\Delta\chi^2$ is zero. The black lines are the $1\sigma$ contour taken from the recent NO$\nu$A results~\cite{NOvA:2021nfi}, recent T2K results~\cite{T2K:2023smv} and recent joint result of T2K and NO$\nu$A~\cite{T2K:2025wet}, respectively. The black solid dots represent the best-fit values as given by the collaborations~\cite{T2K:2025wet}.}} 
	\label{fig:Contour}
\end{figure}

We address the signature of LMS scheme in the correlations between $\sin^2\theta_{23}$ and the third most uncertain oscillation parameters, $\textit{i.e.,}$ $\Delta m^2_{31}$ in the light of NO$\nu$A and T2K in this section through figure~\ref{fig:Contour}, considering true NMO. Here also, we kept the choice of the benchmark oscillation parameters (except $\theta_{23}$ and $\Delta m^2_{31}$) at the local minima of total concurrence as per the aim to keep both the experiments in the same footing [see table~\ref{tab:NMO_concurrence_combined}]. Figure~\ref {fig:Contour} depicts the influence of minimum concurrence in the allowed region in ($\sin^2\theta_{23}-\Delta m^2_{31}$) parameter space at $1\sigma$ C.L. in NO$\nu$A (left panel), T2K (middle panel), and the combined setup of NO$\nu$A and T2K (right panel), respectively. The upper (lower) panels are for the benchmark choices according to set-1; blue lines (set-2; green lines) corresponding to the local minima of total concurrence as given in table~\ref{tab:NMO_concurrence_combined}. The colored triangular dots represent the coordinates of this parameter space for which $\Delta \chi^2=0$. On the other hand, the black lines and the black solid dots represent the $1\sigma$ favorable region and the best-fit values in this parameter space as given by the experimental collaborations~\cite{T2K:2025wet}. In figure~\ref{fig:Contour}, our $\Delta\chi^2$ is defined as follows,\\
\begin{equation}
\begin{aligned}
 \Delta \chi^2 = \underset{(\kappa_s,\,\kappa_b,\,\delta_\mathrm{CP})}{\mathrm{min}} \bigg[
    &\chi^2(\sin^2\theta_{23}^{\mathrm{test}} \in [0.4:0.6], \, \Delta m^2_{31} \in [2.450:\,2.578]\times 10^{-3}\,\mathrm{eV}^2) \\
    &-\chi^2(\sin^2\theta_{23}^{\mathrm{true}},\,\Delta m^{2,{\mathrm{true}}}_{31} \in \mathrm{Real\,\, Data})\bigg].
\end{aligned}
\label{eq:chi2-sensitivity-contour-real_th23_ldm}
\end{equation}

\begin{figure}[htb!]
	\centering 
 \includegraphics[width=\linewidth]{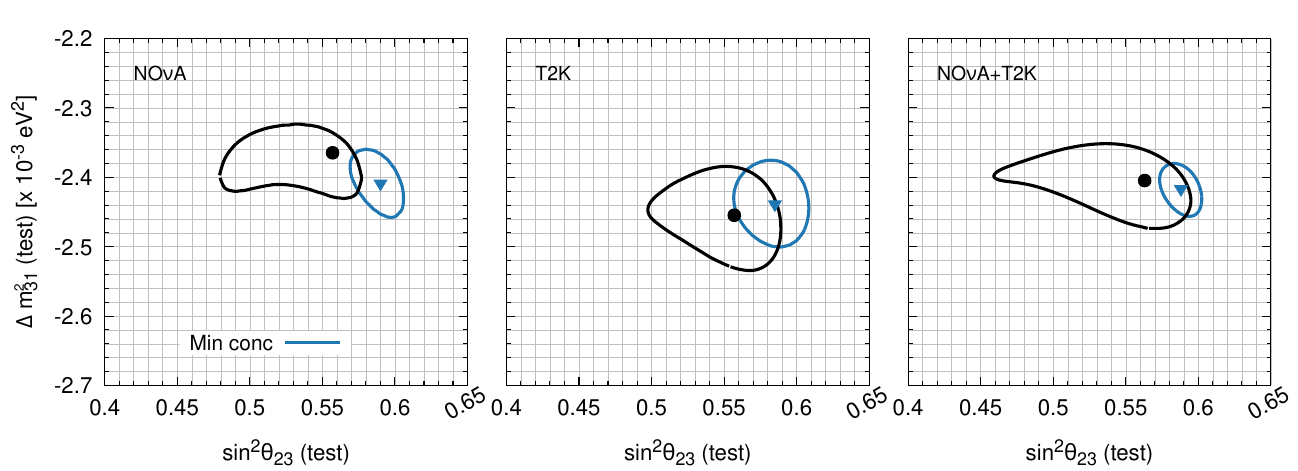}
	\caption{\footnotesize{Same as figure~\ref{fig:Contour} using the common oscillation parameter set given in table~\ref{tab:IMO_concurrence_combined} for true IMO. 
    }} 
	\label{fig:imo-mass}
\end{figure}

We marginalize our $\chi^2$ over the $3\sigma$ uncertainty of $\delta_{CP}$, as mentioned in table~\ref{table:NuFit}, alongside the systematic uncertainties of the signal ($\kappa_s$) and the background ($\kappa_b$). As illustrated in figure~\ref{fig:Contour}, a significantly tighter $1\sigma$ allowed region for NO$\nu$A (indicated by both blue and green contours) relative to the official experimental bounds (black contours) is observed, whereas T2K exhibits only a marginal constraint. This behavior is elucidated by figure~\ref{fig:2_conc_vs_energy}: under the LMS scheme, the local minima of total concurrence shift significantly closer to the energy range, corresponding to the peak flux of NO$\nu$A, while T2K experiences a negligible shift. Because the T2K beam configuration naturally aligns its peak flux near the local concurrence minima, the resulting parameter space confinement is less pronounced than in NO$\nu$A. Furthermore, NO$\nu$A's larger detector volume enhances constraints on $\sin^2\theta_{23}$, while its longer baseline induces stronger matter effects that yield a more precise determination of $\Delta m^2_{31}$ compared to T2K, consistent with official collaboration reports. Ultimately, a joint analysis leverages both higher statistics and the complementary features of these experiments to produce more stringent bounds. Crucially, while neither independent nor combined standard experimental setups can break the $\theta_{23}$ octant degeneracy, selecting oscillation parameters that shifts local minima of total concurrence within FWHM, significantly increases the number of events, and hence expounds more precision for both $\theta_{23}$ and $\Delta m^2_{31}$ and resolve the $\theta_{23}$ octant ambiguity, even in the individual setups. The best-fit points, indicated by colored triangles, demonstrate good agreement with the experimental best-fit values from individual NO$\nu$A and T2K data, as well as the joint NO$\nu$A+T2K fit. We find that, in the presence of minimum total concurrence corresponding to the benchmarks of set-1, the co-ordinates of the best-fit in ($\sin^2\theta_{23}-\Delta m^2_{31}$) plane are ($0.581^{+0.0140}_{-0.0150},\,{2.522^{+0.0364}_{-0.0348}}\times10^{-3}\,\mathrm{eV}^2$), and for set-2, they are ($0.580^{+0.0140}_{-0.0153},\,{2.515^{+0.0344}_{-0.0344}}\times10^{-3}\,\mathrm{eV}^2$); whereas the joint collaboration of NO$\nu$A and T2K~\cite{T2K:2025wet} gives their best-fit at ($0.561^{+0.021}_{-0.039},\,{2.505^{+0.1153}_{-0.1053}}\times10^{-3}\,\mathrm{eV}^2$).

We perform the same analysis for the true IMO using the oscillation parameters listed in Table \ref{tab:IMO_concurrence_combined}, and the corresponding results are presented in figure~\ref{fig:imo-mass}.
Unlike figure~\ref{fig:Contour}, figure~\ref{fig:imo-mass} consists of only three panels, as our LMS scheme yields a single set of common oscillation parameters for this scenario. The observations and their interpretation are consistent with those discussed for figure~\ref{fig:Contour}. As shown in figure ~\ref{fig:imo-mass}, our analysis consistently favors the higher octant in all three panels, thereby resolving the $\theta_{23}$ octant degeneracy.
We find that, in the presence of minimum total concurrence corresponding to the benchmarks of table~\ref{tab:IMO_concurrence_combined}, the co-ordinates of the best-fit in ($\sin^2\theta_{23}-\Delta m^2_{31}$) plane are ($0.588^{+0.0135}_{-0.0134},\,{-2.418^{+0.0359}_{-0.0360}}\times10^{-3}\,\mathrm{eV}^2$) for IMO; whereas the joint collaboration of NO$\nu$A and T2K~\cite{T2K:2025wet} gives their best-fit at ($0.563^{+0.021}_{-0.039},\,{-2.405^{+0.0353}_{-0.0453}}\times10^{-3}\,\mathrm{eV}^2$).


\subsection{Effect of LMS scheme on examining CP violation}
\begin{figure}[htb!]
	\centering 
\includegraphics[width=0.98\linewidth]{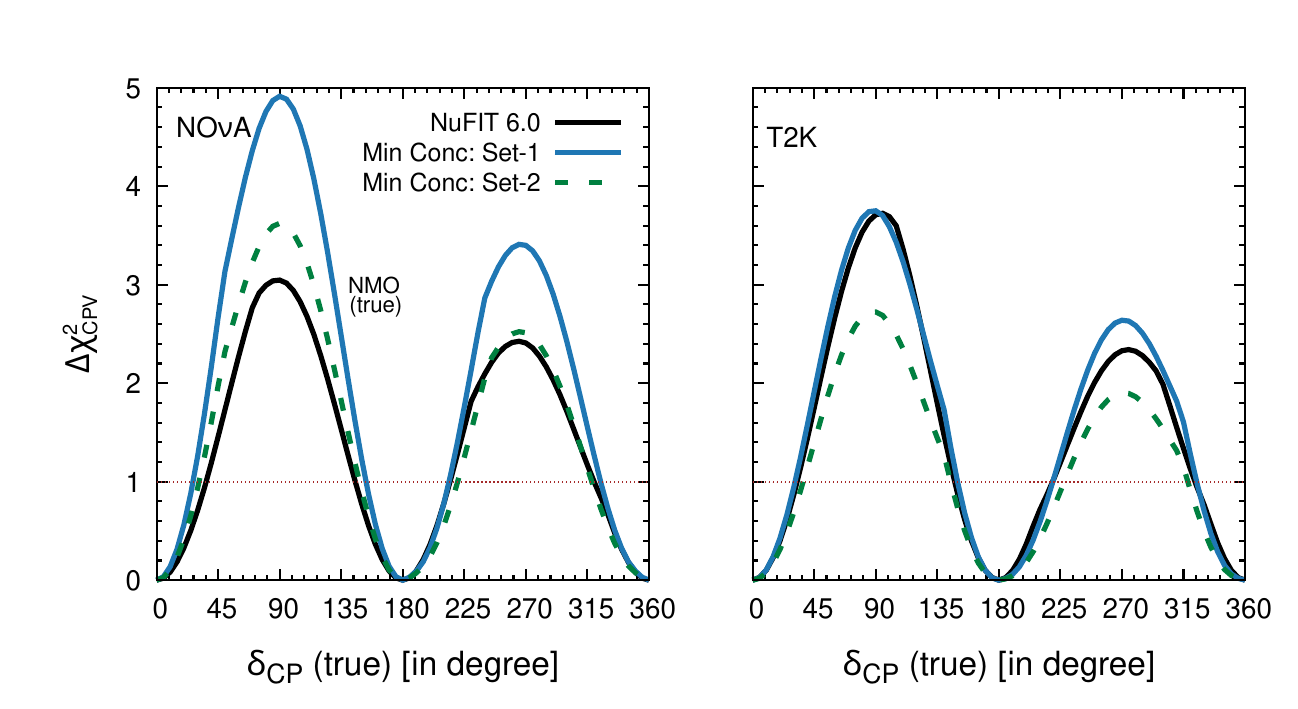}
	\caption{\footnotesize{The figure displays the effect of the LMS scheme in probing CP violation in the neutrino sector at NO$\nu$A and T2K.  Blue-solid and green-dashed lines show the sensitivity of NO$\nu$A (left) and T2K (right) for set-1 and set-2 concurrence minima, respectively, of table~\ref{tab:NMO_concurrence_combined}, whereas the black line denotes the same for the standard oscillation parameters from NuFit 6.0 (see table~\ref{table:NuFit}). Both plots represent the sensitivity to CPV as a function of true $\delta_\mathrm{CP}$. Here, marginalization has been performed over $\delta_{\mathrm{CP}}$ and $\Delta m^2_{31}$ into their 3$\sigma$ span.
    }} 
	\label{fig:CPV_with_max_min_conc}
\end{figure}
Neutrino oscillation studies are a good handle for probing leptonic CP violation, a highly sought-after quest to answer the question of matter-antimatter asymmetry~\cite{Pascoli:2006ci, T2K:2019bcf, Rahaman:2022rfp, Agarwalla:2013avs, Ghosh:2013yon}. Here, we illustrate the effect of LMS scheme on the capability of detecting CP violation, assuming neutrino as a Dirac fermion. In the left (right) column of figure~\ref{fig:CPV_with_max_min_conc}, we show the influence of local minima of total concurrence by simulating the ongoing NO$\nu$A (T2K) experiment according to our goal of study. For this, the quantity $\Delta\chi^2_\mathrm{CPV}$ is defined as follows.\\

\begin{equation}
 \Delta \chi^2_\mathrm{CPV} = \underset{( \sin^2\theta_{23},\,\Delta m^{2}_{31},\,\kappa_s,\,\kappa_b)}{\mathrm{min}} \bigg[
    \chi^2(\delta_{\mathrm{CP}}^{\mathrm{test}} \in \{0^\circ,\,180^\circ,\,360^\circ\})-\chi^2(\delta_{\mathrm{CP}}^{\mathrm{true}} \in [0^{\circ}, 360^{\circ}])\bigg].
    \label{eq:chi2-sensitivity-cpv}
\end{equation}
Here, we marginalize the $\Delta\chi^2_\mathrm{CPV}$ over the $3\sigma$ range of $\sin^2\theta_{23},$ and $\Delta m^{2}_{31},$ as mentioned in~\cite{Esteban:2024eli} 
after considering the effects of systematic uncertainties of signals and backgrounds, $\,\kappa_s$ and $\kappa_b$, respectively. We fix our theory at CP-conserving angles $0^\circ$, $180^\circ$, and $360^\circ$ and varied the true value of $\delta_\mathrm{CP}$ for all possible values (\textit{i.e.,} $0^\circ$ to $360^\circ$). The benchmark choices of the green and blue lines are according to the set 1 and set 2 concurrence minima, respectively, in table~\ref{tab:NMO_concurrence_combined}. The black line depicts the sensitivity of CP violation, taking the benchmark choices as mentioned in the global-fit~\cite{Esteban:2024eli} or in table~\ref{table:NuFit}.
We consider the true mass ordering as NMO for the analysis for figure~\ref{fig:CPV_with_max_min_conc}.

In figure~\ref{fig:CPV_with_max_min_conc}, we observe that NO$\nu$A is more capable of establishing CP violation at true $\delta_\mathrm{CP}= 90^\circ, 270^\circ$ for both set 1 and set 2 given in table~\ref{tab:NMO_concurrence_combined}, rather it outperforms the sensitivity for the NuFit 6.0 global fit parameters. On the other hand, for T2K, the capability of examining the CP-violation is increased for true $\delta_\mathrm{CP}=270^\circ$ for the set 1 benchmark values of table~\ref{tab:NMO_concurrence_combined}. However, the sensitivity to CP-violation significantly decreases for benchmark values given in set 2.

CP-violation sensitivity depends on the distinguishability of the CP-violating event spectrum from the CP-conserving hypotheses, $\delta_{CP}=0,\pi$, over the full reconstructed-energy distribution. Therefore, moving the concurrence minimum closer to the flux peak may also reduce CPV sensitivity if the shifted spectrum becomes more similar to one of the CP-conserving spectra, or if the spectral separation is redistributed away from energy bins that were previously more discriminating. In this case, the local statistical weight improves, but the global CPV hypothesis separation can become weaker. 
\subsection{Effect of LMS scheme in eliminating the wrong $\theta_{23}$ octant}
\begin{figure}[htb!]
	\centering 
 \includegraphics[width=0.98\linewidth]{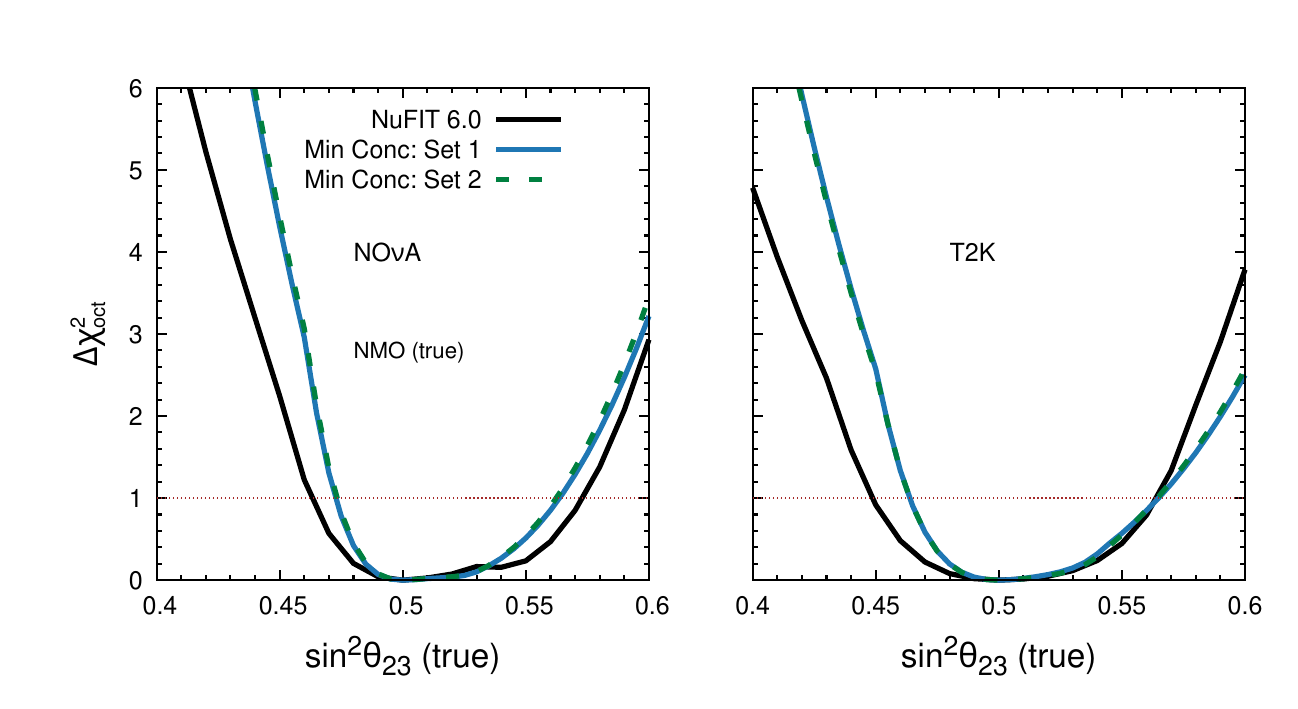}
 \caption{\footnotesize{The figure displays the effect of LMS scheme in excluding the wrong solution of $\theta_{23}$ octant at NO$\nu$A and T2K. Blue-solid and green-dashed lines show the sensitivity of NO$\nu$A (left) and T2K (right) for set-1 and set-2 concurrence minima (see table~\ref{tab:NMO_concurrence_combined}), respectively, whereas the black line denotes the same for the standard oscillation parameters from NuFit 6.0 (see table~\ref{table:NuFit}). Both plots show the sensitivity to the exclusion of the opposite $\theta_{23}$ octant, depicted as a function of true $\sin^2\theta_{23}$. Here, marginalization has been performed over $\delta_{\mathrm{CP}}$ and $\Delta m^2_{31}$ into their 3$\sigma$ span.}}
	\label{fig:Oct_excl_with_max_min_conc}
\end{figure}
Determining the correct $\theta_{23}$ octant is still an open-ended issue in the neutrino oscillation domain. In the neutrino sector, $\theta_{23}$ is the second most uncertain parameter after $\delta_{\mathrm{CP}}$~\cite{Agarwalla:2013ju, Agarwalla:2021bzs, Ghosh:2015ena}, hence it is important to settle down the correct octant of $\theta_{23}$. In this section, we see how our LMS scheme affects the exclusion of the opposite $\theta_{23}$ octant in the present long-baseline experiments NO$\nu$A and T2K and present our results in figure~\ref{fig:Oct_excl_with_max_min_conc}. For this purpose, $\Delta\chi^2_{\mathrm{oct}}$ is defined as follows
\begin{eqnarray}
    \Delta \chi^{2}_{\text{oct}} &=& \underset{(\vec{\lambda})}{\mathrm{min}}\left\{ \chi^2\left(\sin^2\theta_{23}^{\mathrm{test}} \right) - \chi^2\left(\sin^2\theta_{23}^{\mathrm{true}}\right)\right\}\,.
\label{eq:octant-exclusion-chi2}
\end{eqnarray}
Here, while our hypothesis is that $\theta_{23}$ lies in lower octant (LO), \textit{i.e.,} $\sin^2\theta_{23}^{\mathrm{test}}\in [0.4,0.5]$, the true value of $\theta_{23}$ is in higher octant (HO), \textit{i.e.,} $\sin^2\theta_{23}^{\mathrm{true}}\in [0.5,0.6]$. But, while the hypothesis is $\theta_{23}$ in HO, \textit{i.e.,} $\sin^2\theta_{23}^{\mathrm{test}}\in [0.5,0.6]$, the true value of $\theta_{23}$ is in LO, \textit{i.e.,} $\sin^2\theta_{23}^{\mathrm{true}}\in [0.4,0.5]$. So, we are excluding the opposite $\theta_{23}$ octant in this analysis as we do not know which octant is true in Nature. Here, $\lambda$ is the marginalized parameter for evaluating $\Delta\chi^2_\mathrm{oct}$ and we take $\lambda=\delta_\mathrm{CP},\,\Delta m^2_{31}$ (in their $3\sigma$ range as given in table~\ref{table:NuFit}), $\kappa_s,$ and $\kappa_b$. Here, the benchmark choices of the oscillation parameters for the blue (green) line are taken from set 1 (set 2) of table~\ref{tab:NMO_concurrence_combined}. 

 In figure~\ref{fig:Oct_excl_with_max_min_conc}, we observe that the $\theta_{23}$ octant exclusion sensitivity of NO$\nu$A, for benchmark values of set 1 and set 2 in table~\ref{tab:NMO_concurrence_combined} (blue and green lines) outperforms the standard case (black line, where we use benchmarks from NuFIT 6.0) in case of both true LO and HO.  
 We also observed similar results for T2K, except for a very small range in higher octant where a decrease in the sensitivity can be seen for both blue and green curves. The octant sensitivity is a global degeneracy-resolution problem rather than a purely local precision measure. It depends on the separation between the true-octant spectrum and the best-fit wrong-octant spectrum after marginalization over $\delta_{CP}$ and $\Delta m_{31}^2$, and systematic uncertainties. In such cases, the statistical gain from the shift may be offset by increased octant degeneracy, resulting in reduced octant sensitivity.


\subsection{Effect of LMS scheme in examining IMO exclusion sensitivity}
\begin{figure}[H]
	\centering 
\includegraphics[width=0.98\linewidth]{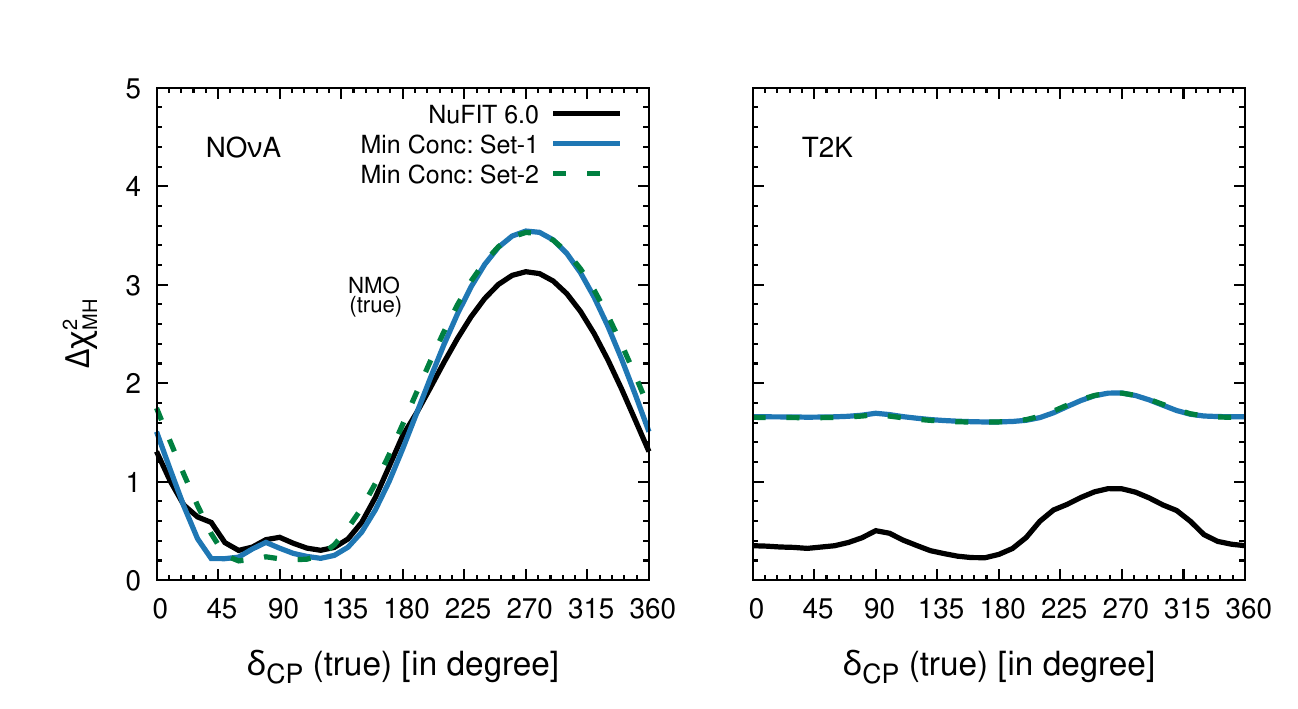}
\caption{\footnotesize{The figure displays the effect of LMS scheme in probing IMO exclusion in the neutrino sector at NO$\nu$A and T2K.  Blue-solid and green-dashed lines show the sensitivity of NO$\nu$A (left) and T2K (right) for set-1 and set-2 concurrence minima (see table~\ref{tab:NMO_concurrence_combined}), respectively, whereas the black line denotes the same for the standard oscillation parameters from NuFit 6.0 (see table~\ref{table:NuFit}). Both plots represent the sensitivity to IMO exclusion as a function of true $\delta_{CP}$. Here, marginalization has been performed over $\theta_{23}$ into its 3$\sigma$ span.}} 
	\label{fig:IMO_excl_sensitivity_with_max_min_conc}
\end{figure}

We expound the influence of LMS scheme for minimum concurrence on the sensitivity of excluding the IMO in the light of NO$\nu$A and T2K. Although the present global fit data slightly favors the NMO with $2.5\sigma$ preference~\cite{Esteban:2024eli}, neutrino mass-hierarchy ambiguity still cannot be completely resolved either in the ongoing~\cite{Dixit:2018kev, Petcov:2005rv, Agarwalla:2013qfa, Prakash:2013nra} or in the future long baseline experiments~\cite{Sarker:2023qzp, Sarker:2026zqv} till date. In figure~\ref{fig:IMO_excl_sensitivity_with_max_min_conc}, we assume NMO as the true neutrino mass ordering, whereas IMO is the mass ordering in the test spectrum, however this sensitivity is displayed as the function of true $\delta_{CP}$ in its full possible range, as it is the most uncertain oscillation parameter. The considered uncertainty ranges of the parameters are as per the table~\ref{table:NuFit}. The benchmark choices for the blue (green) lines are given by set-1 (set-2) of the minimum total concurrence as shown in table~\ref{tab:NMO_concurrence_combined},  whereas the black lines take the benchmarks as given in NuFIT 6.0 in table~\ref{table:NuFit}. Here, the definition of $\Delta\chi^2_\mathrm{MH}$ is as follows,\\
\begin{equation}
\begin{aligned}
 \Delta \chi^2_\mathrm{MH} = \underset{( \sin^2\theta_{23},\,\kappa_s,\,\kappa_b)}{\mathrm{min}} \bigg[
    \chi^2\big(\Delta m^{2,\mathrm{test}}_{31} \in [-2.55:-2.42]\times10^{-3}\,\mathrm{eV}^2\big) \\
    {}- \chi^2\big(\Delta m^{2,\mathrm{true}}_{31} \in [2.45:2.58]\times10^{-3}\,\mathrm{eV}^2\big)\bigg],
\end{aligned}
\label{eq:chi2-sensitivity-MH}
\end{equation}
where, we marginalize our $\chi^2$ in the $3\sigma$ range of $\sin^2\theta_{23}$ and the systemetic uncertainties of signal ($\kappa_s$) and the background $\kappa_b$. We consider the variation of $\Delta m^2_{31}$ in their $3\sigma$ range as given in NuFIT 6.0~\cite{Esteban:2024eli} for both true (NMO) and test (IMO) values.\\

In figure~\ref{fig:IMO_excl_sensitivity_with_max_min_conc}, we investigate that, both NO$\nu$A and T2K can exclude IMO more in the presence of minimum concurrence than the standard case, where we use benchmarks from NuFIT 6.0 at true $\delta_\mathrm{CP}=270^\circ$. But, in the context of the inter-competition between the two experiments at true $\delta_\mathrm{CP}=270^\circ$, NO$\nu$A outperforms T2K, either in the standard case,where using benchmarks from NuFIT 6.0 or in the presence of minimum concurrence. NO$\nu$A's enhanced matter effect due to its longer baseline helps it to achieve more sensitivity to our study. However, both sets of benchmark oscillation parameters corresponding to minimum concurrences show almost equal sensitivity to exclude IMO. At maximum CP-violating phase ($\textit{i.e.,}\,\mathrm{true}\,\delta_\mathrm{CP}=270^\circ$), T2K shows more significant enhancement than the standard case while taking benchmark choices from NuFIT 6.0, compared to NO$\nu$A. This happens as the local minima of the concurrence does not shift much for T2K from its 1st oscillation maxima (at $E=0.58$ GeV), and that energy bin also contributes more in the IMO exclusion sensitivity than the other energy bins, which is not happening in NO$\nu$A.

\section{Discussion and Conclusions}
 \label{sec:Discussion}
The long baseline NO$\nu$A and T2K experiments have had tensions in their results, specifically on the value of the CP-violating phase and the $\theta_{23}$ octant. Many previous analyses have attempted to resolve this tension using new physics effects, such as non-standard interactions~\cite{Chatterjee:2020kkm, Denton:2020uda}, non-unitarity of the mixing matrix~\cite{Miranda:2019ynh}, Lorentz invariance violation~\cite{Rahaman:2021leu} and decoherence effects~\cite{Coelho:2017zes, Stankevich:2018wgz, Loulijat:2026vyq} and the tension can be alleviated to varying degrees, see~\cite{Rahaman:2021zzm} for review. In this work, we analyze the results of NO$\nu$A and T2K through the entanglement properties of the flavor state, with particular emphasis on the local minima of the entanglement measure.

We investigated how local minima of total concurrence, an entanglement measure, affects the extraction of a preferred set of oscillation parameters in the long-baseline experiments NO$\nu$A and T2K. These minima identify energy windows in which the flavor state is closest to separability, potentially enabling a cleaner determination of oscillation parameters. We show that the local minima of total concurrence closely coincide with maxima of QFI. Since QFI plays a central role in quantum metrology and sets the lower bound on parameter uncertainty through an inverse-square-root scaling, $\Delta \alpha \propto 1/\sqrt{H_Q(\alpha)}$, its maxima corresponds to the minimum achievable uncertainty in parameter measurement. It supports the argument for clear measurement near local minima of entanglement, where the flavor state is nearly separable. 

Further, to minimize the tension arising from differing energy regimes between NO$\nu$A and T2K, we choose the common oscillation parameter sets that minimize the concurrence for both experiments within their respective FWHM around the peak fluxes and use these sets as benchmark parameter values. Consequently, this procedure of choosing benchmark values corresponding to the minimum concurrence shifts the local minima of total concurrence and the maxima of QFI closer to the energy values associated with the peak flux for each experiment. We call it the LMS scheme within the higher event statistics region. 

Finally, we present the results analyzing the effects of this LMS scheme on the extraction of various oscillation parameters, specifically, CP-violating phase $\delta_{CP}$, mixing angle $\theta_{23}$ and mass squared difference $\Delta m^2_{31}$. We observed that the uncertainty in these three parameters is significantly reduced. Important point to note is that both experiments favor the higher octant ($>45^\circ$) of $\theta_{23}$, irrespective of the type of mass ordering. However, we also observed that NMO is favored over IMO by both experiments. The numbers we observed are $\delta_{CP} = 195^{+38}_{-32}\,^\circ$ (for NMO), $\delta_{CP} = 294^{+29.9}_{-38.2}\,^\circ$ (for IMO), $\sin^2\theta_{23} = 0.581^{+0.0136}_{-0.0150}$ (for NMO), $\sin^2\theta_{23} = 0.595^{+0.0115}_{-0.0126}$ (for IMO) and $\Delta m^2_{31} = 2.515^{+0.0344}_{-0.0344}\times10^{-3}\,\mathrm{eV}^2$ (for NMO), $\Delta m_{31}^2 = {-2.418^{+0.0359}_{-0.0360}}\times10^{-3}\,\mathrm{eV}^2$ (for IMO).

 We also examine the effects of the LMS scheme on the sensitivity of both experiments to examine CP-violation, to determine the mass ordering, and to eliminate the wrong $\theta_{23}$ octant. Interestingly, we observe a significant increase in the sensitivity to all these three parameters compared to the NuFIT 6.0 global best-fit parameter set in case of NO$\nu$A experiment. In case of T2K, we observe a significant enhancement in the sensitivity to the IMO exclusion sensitivity. We also observed a significant increase in sensitivity to eliminate the wrong $\theta_{23}$ octant, except for some $\sin^2\theta_{23}$ values between 0.57 and 0.6, where a slight reduction in sensitivity is observed for T2K. In the context of the CP-violation, a small increase in sensitivity is observed for set-1 of common oscillation parameters corresponding to the minimum concurrence, while a significant decrease for set-2 is observed. The LMS scheme can improve the precision of continuously varying oscillation parameters, since small parameter changes are then probed in an energy region with higher event statistics. However, CP-violation sensitivity is not determined solely by the precision of local parameters. It depends on the distinguishability of the CP-violating event spectrum from the CP-conserving hypotheses, over the full reconstructed-energy distribution. Therefore, moving the concurrence minimum closer to the flux peak may also reduce CP-violation sensitivity if the shifted spectrum becomes more similar to one of the CP-conserving spectra, or if the spectral separation is redistributed away from energy bins that were previously more discriminating. In this case, the local statistical weight improves, but the global CPV hypothesis separation can become weaker.

 Recent works have demonstrated that QFI provides a powerful quantum-metrological diagnostic for neutrino oscillations, yielding measurement-independent precision bounds and clarifying how efficiently flavor measurements extract the information encoded in the neutrino state \cite{Ignoti:2025rxr, Frugiuele:2026yeq, Chundawat:2026lcm, Huang:2026bws, Yadav:2026mnw, Lu:2026wzr}. These analyses primarily focus on the QFI content itself, the associated quantum Cramér-Rao bounds, and the comparison between intrinsic and experimentally accessible Fisher information. In the present work we go beyond this information-content analysis and develop a complementary direction by relating parameter sensitivity to the behavior of quantum correlations, specifically the local minima of total concurrence. We show that these minima can be exploited not only as diagnostic markers of favorable energy windows, but also as a practical guide for improving the extraction of oscillation parameters in existing long-baseline experiments. Applying this strategy to NO$\nu$A and T2K, we study its effect on joint parameter constraints and on key physics sensitivities, including leptonic CP violation, $\theta_{23}$ octant determination, and establishment of the correct mass ordering, while also examining its role in reducing the tension between the two experiments within their present experimental limitations.

 To our knowledge, this work presents the first application of entanglement-minimum-based optimization to an analysis incorporating\textit{ real long-baseline neutrino oscillation data} from T2K and NO$\nu$A. Banerjee et al. previously studied concurrence minima in a simulated DUNE setup and demonstrated their relevance for parameter sensitivity in \cite{Banerjee:2024lih}. Our study is also the first to connect local minima of an entanglement measure with maxima of QFI, which directly characterizes the ultimate precision attainable in parameter estimation. A related study by Nogueira et al. linked von Neumann entropy-based entanglement with the \textit{classical} Fisher information in a two-flavor scenario; however, their QFI was constant, and no correspondence between entanglement minima and QFI maxima was found \cite{Nogueira:2016qsk}.

 The central physical conclusion of the present work concerns the identification of energy regions where the flavor state approaches the local minimum of the entanglement measure and where the QFI is simultaneously enhanced. We also verified our results for the genuine tripartite entanglement measure, for example, three-$\pi$ ($\pi_{\mu e\tau}$) \cite{Jha:2020dav}, and observed that the energy values corresponding to the local minima of $\pi_{\mu e\tau}$ and $C^{\mu^2}$ coincide with each other. This finding also demonstrates the robustness of our results in the context of genuine tripartite entanglement.

 For long baseline experiments, such as T2K and NO$\nu$A, these concurrence minima also coincide with the minima of the survival probability, $P_{\nu_\mu\rightarrow\nu_\mu}$, and maxima of $P_{\nu_\mu\rightarrow\nu_\tau}$, while $P_{\nu_\mu\rightarrow\nu_e}$ remains comparatively suppressed because of the small value of $\theta_{13}$.  The flavor state is therefore close to separable in these regions, being dominated by the $\nu_\tau$ component \cite{Banerjee:2024lih}. Thus, similar ``sweet spots'', as shown in our analysis in terms of set-1 and set-2 under the LMS scheme, could also be inferred from extrema of the oscillation probabilities. However, such probability extrema do not by themselves guarantee optimal parameter precision. Their significance is established by their correspondence to the concurrence minima and, more importantly, to the QFI maxima, which directly quantify the ultimate precision achievable in estimating the oscillation parameters.

 As a final remark, we note that T2K and NO$\nu$A probe different energy ranges and baselines, and therefore sample regions with different intrinsic information content on the oscillation parameters. The T2K experiment has been observed to have its flux peak coinciding with local minima of total concurrence and maxima of QFI at a common energy. In contrast, for the NO$\nu$A experiment, the FWHM region around neutrino flux peak energy is located away from the local minima of concurrence and the maxima of QFI. This is indicated in Fig 1 of our manuscript. Since the QFI sets the quantum Cramér–Rao bound on achievable precision, parameter extraction becomes intrinsically less precise in energy regions where the QFI is smaller, irrespective of the statistical analysis employed. In addition, the event statistics can also play an important role to obtain the ultimate precision. We therefore do not claim that the T2K–NO$\nu$A tension is solely a consequence of the ultimate quantum precision limit achievable, but our results suggest that the differing proximity of the two experiments to high QFI energy regions may contribute to the different preferred parameter regions. It still leaves space for new physics effects to further reduce the tension as explored in Refs.~[44,104-110] cited in our paper.

This work highlights the role of quantum information measures as useful diagnostic tools for understanding where oscillation parameters are intrinsically most accessible in neutrino flavor evolution. When combined with realistic flux, statistics, and spectral distinguishability, these tools can provide deeper insight into both precision estimation and sensitivity studies for CP violation, mass ordering, and parameter degeneracies.

\acknowledgments  
We especially thank Ushak Rahaman for his helpful discussions and contributions to the GLoBES simulation framework. K.D. and S.R. would like to acknowledge support from the National Research Foundation of South Africa through a NITheCS grant. The numerical simulations were performed using the SAMKHYA: High-Performance Computing Facility at the Institute of Physics, Bhubaneswar. RK acknowledges the help of Makrand Siddhabhatti in accessing the HPC SAMKHYA.



\appendix
\label{Appendix}
\renewcommand\thefigure{A\arabic{figure}}
\renewcommand\theHfigure{A\arabic{figure}}
\renewcommand\thetable{A\arabic{table}}
\renewcommand\theHtable{A\arabic{table}}
\setcounter{table}{0} 
\setcounter{figure}{0} 
\setcounter{equation}{0}


\bibliographystyle{JHEP}
\bibliography{nuqi}
\end{document}